\documentclass[aps, prd, amsmath, floats, twocolumn, floatfix, nofootinbib, superscriptaddress]{revtex4-1}
\usepackage{graphicx}
\usepackage{xcolor}
\usepackage{latexsym}
\usepackage{epsfig}
\usepackage[colorlinks]{hyperref}
\usepackage{array}
\usepackage{enumitem}
\usepackage{textcomp}
\usepackage{footmisc}
\setlist{noitemsep,leftmargin=*}

\newcommand{\be}{\begin{eqnarray}}
\newcommand{\ee}{\end{eqnarray}}
\newcommand{\bg}{\begin{gather}}
\newcommand{\eg}{\end{gather}}
\newcommand{\bead}{\begin{eqnarray}\begin{aligned}}
\newcommand{\eead}{\end{aligned}\end{eqnarray}}
\newcommand{\beq}{\begin{equation*}}
\newcommand{\eeq}{\end{equation*}}

\newcommand{\bitem}{\begin{itemize}}
\newcommand{\eitem}{\end{itemize}}
\newcommand{\bwide}{\begin{widetext}}
\newcommand{\ewide}{\end{widetext}}

\newcommand{\slsim}{\, \scriptsize{\lesssim}\,}

\begin{document}

\title{Shining X-rays on asymptotically safe quantum gravity}

\author{Biao~Zhou}
\affiliation{Center for Field Theory and Particle Physics and Department of Physics, Fudan University, 200438 Shanghai, China}

\author{Askar~B.~Abdikamalov}
\affiliation{Center for Field Theory and Particle Physics and Department of Physics, Fudan University, 200438 Shanghai, China}

\author{Dimitry~Ayzenberg}
\affiliation{Center for Field Theory and Particle Physics and Department of Physics, Fudan University, 200438 Shanghai, China}
\affiliation{Theoretical Astrophysics, Eberhard-Karls Universit\"at T\"ubingen, 72076 T\"ubingen, Germany}

\author{Cosimo~Bambi}
\email[Corresponding author: ]{bambi@fudan.edu.cn}
\affiliation{Center for Field Theory and Particle Physics and Department of Physics, Fudan University, 200438 Shanghai, China}
 
\author{Sourabh~Nampalliwar}
\affiliation{Theoretical Astrophysics, Eberhard-Karls Universit\"at T\"ubingen, 72076 T\"ubingen, Germany}

\author{Ashutosh~Tripathi}
\affiliation{Center for Field Theory and Particle Physics and Department of Physics, Fudan University, 200438 Shanghai, China}

\begin{abstract}
Asymptotically safe quantum gravity is a promising candidate scenario to provide a UV extension for the effective quantum field theory of Einstein's gravity. The theory has its foundations on the very successful framework of quantum field theory, which has been extensively tested for electromagnetic and nuclear interactions. However, observational tests of asymptotically safe quantum gravity are more challenging. Recently, a rotating black hole metric inspired by asymptotically safe quantum gravity has been proposed, and this opens the possibility of astrophysical tests of the theory. In the present paper, we show the capabilities of X-ray reflection spectroscopy to constrain the inverse dimensionless fixed-point value $\gamma$ from the analysis of a \textsl{Suzaku} observation of the X-ray binary GRS~1915+105. We compare these constraints with those obtained from black hole imaging.
\end{abstract}

\maketitle


\section{Introduction \label{s-intro}}

Einstein's theory of gravity, known as general relativity, is the standard theory to describe gravitational phenomena in our Universe. Since it was proposed back in 1915, a large variety of experiments have been performed to check its veracity. For most of history, these tests were able to probe only the weak-field regime~\cite{will2014}, and general relativity (GR hereafter) proved to be highly successful. The story in the strong-field regime is quite interesting. Over the past few years, tests in the strong-field regime have become commonplace~\cite{Bambi2015,LIGOScientific:2019fpa,Carson:2019rda,Vincent:2020dij}. Though early results are in full agreement with GR~\cite{LIGOScientific:2019fpa,Akiyama:2019eap,Abdikamalov:2019zfz}, theoretical considerations cast doubt on the veracity of GR in all regimes.
These include the presence of singularities~\cite{Penrose:1964wq}, and the difficulties to find a theory of quantum gravity beyond an effective low-energy model~\cite{tHooft:1974toh}. Both these issues suggest that there is some theory that supersedes GR, and of which GR is the low energy limit. Various proposals of such superseding theories are matters of active research, the most popular ones being string theory~\cite{Aharony:1999ti} and loop quantum gravity~\cite{Ashtekar:2011ni}. 
Determining the validity of any such proposal is impossible if the theory does not have any testable predictions.

Na\"{i}vely merging GR with quantum mechanics results in a failure, due to the unfortunate fact that  GR is nonrenormalizable~\cite{tHooft:1974toh,vandeVen:1991gw}. Workarounds have been found to resolve this issue, one of them being the idea of asymptotic safety~\cite{Bonanno:2000ep,Reuter:2012id,Bonanno:2020bil}. {The basic idea of asymptotically safe gravity, advocated by Weinberg back in 1996~\cite{Weinberg:1996kw},} is the following~\cite{Weinberg:1980gg,Held:2019xde}: quantum fluctuations modify the standard gravitational interactions by making them scale dependent. This can be achieved by turning the Newton's coupling constant $G_N$ into a length dependent constant $G_N(r)$. But such a modification can result in infinities in the theory, which is a deal-breaker. In typical quantum field gravity, this problem is resolved by invoking asymptotic freedom, which prescribes that particle interactions become asymptotically weaker as the corresponding length scale decreases. But in GR modified with a scale dependent $G(r)$, perturbative renormalization breaks down, rendering asymptotic freedom impossible~\cite{tHooft:1974toh,vandeVen:1991gw}. {One resolution of this issue is asymptotically safe gravity}. In {an asymptotically safe} setting, the dimensionless coupling parameter ($g(k) \equiv G_N(k)\,k^2$, where $k$ is the Renormalization Group energy scale) becomes constant below a certain length scale. The above condition can be reformulated as~\cite{Held:2019xde} 
\be
	G_N(k) = \frac{g_{\star}}{k^2},
\ee
where $g_{\star}$ is the asymptotically safe fixed-point value. This results in an effective weakening of gravity above certain energy scales (equivalently, below certain length scales).

Black holes (BHs hereafter) are objects which, owing to their compactness, exhibit the strongest gravitational effects in our Universe. Their simple structure within GR implies that any beyond-GR feature, if imprinted on the BH, will be easily identified. Moreover, they are present in large numbers in the Universe, {with estimates placing $\sim100$ million of them in the Milky Way alone~\cite{Elbert:2017sbr}. All these features make BHs the best probes of testing gravity in our Universe. 
Over the past few years, various different approaches for probing the strong gravity effects around BHs have been realized~\cite{Berti:2019tcy}. Gravitational waves, mostly from binary mergers at present~\cite{LIGOScientific:2019fpa,CardenasAvendano:2019zxd} but also from extreme mass ratio inspirals in future~\cite{Barausse:2020rsu}, are expected to be the best probes of strong gravity (though in some cases, it is expected to be weaker~\cite{Bambi:2013mha,Li:2019lsm} than other techniques). X-ray spectroscopy is a very promising technique~\cite{Reynolds:2013qqa,Bambi2015,Bambi2018} and is quite advanced as compared to some other techniques, but suffers from parameter degeneracies and limited  astrophysical modeling. Imaging of the BH shadow is the latest technique for probing the regions very close to the BH, and although very exciting, is not expected to provide very strong constraints on alternative theories~\cite{Mizuno:2018lxz,Vincent:2020dij}.} 

{Within GR, BHs have at most three free parameters - mass, spin and charge.\footnote{Additional assumptions about regularity of spacetime, asymptotic flatness, etc. go in to this result. Refer to~\cite{Chrusciel2012}, for instance, for a complete list of assumptions.}  Astrophysically, charged BHs are not expected to be common~\cite{Bambi:2008hp}, thus only two parameters, mass and spin, characterize all astrophysical BHs~\cite{Carter1971,Robinson1975}. In asymptotically safe gravity, {the introduction of scale dependence in Newton's coupling constant results in modifications in the BH solutions~\cite{Reuter:2010xb,Koch:2014cqa,Held:2019xde,Platania:2019kyx}}. The BH solution depends on the choice of the non-trivial fixed-point $g_{\star}$ and the metric is modified. Observational signatures of these modified BHs have been analyzed for the first time recently~\cite{Held:2019xde,Rincon:2020iwy}. In~\cite{Held:2019xde}, the authors compare the shadows from regular GR BHs and the modified BHs of asymptotically safe gravity. They find two distinguishing effects of the modifications: there is an overall reduction in shadow size, and additionally, in the case of axisymmetric BHs, a dent in the shadow shape. They use the recent observation of the BH in the center of the $M87$ galaxy~\cite{Akiyama:2019cqa} to constrain the fixed-point value parameter $\gamma$ (defined as the inverse of the asymptotically safe fixed-point value $g_{\star}$) below $2\cdot10^{95}$. Our aim in the present paper is to add to this program of testing for asymptotically safe quantum gravity by analyzing these modified BHs with X-ray reflection spectroscopy.}

{X-ray reflection spectroscopy uses the relativistic smearing of irradiation accretion disks around BHs to learn about the nature of the BHs~\cite{Fabian:1989ej,Laor:1991nc} (See, e.g.,~\cite{Reynolds:2013qqa,Reynolds:2019uxi} for a recent review).} The standard model for analyzing X-ray reflection from astrophysical sources assuming GR BHs, \textsc{relxill}~\cite{Garcia2013, Dauser2014}, has been extended to non-GR BHs with the \textsc{relxill\_nk} suite of models~\cite{relxillnk,Abdikamalov:2019yrr} by some of us. Various non-GR theories~\cite{Bambi:2016wdn,Horowitz:2011cq} and deviation parameters~\cite{Johannsen2015,Konoplya2016} have been tested with this approach~\cite{Nampalliwar:2019iti,Liu:2019vqh,Zhou:2019kwb,Cao2017,Tripathi2018a,Xu2018,Choudhury:2018zmf,Zhou2018a,Zhou2018b,Tripathi:2018lhx,Tripathi:2019bya,Zhang:2019zsn,Zhang:2019ldz,Tripathi:2019fms}, and a public version of the model is available at~\cite{relxillnkweb,relxillnkweb2}. We have implemented the BH solutions of asymptotically safe gravity in the \textsc{relxill\_nk} framework. Furthermore, we have used this framework to analyze data from an X-ray binary in our Galaxy to get astrophysical bounds on the fixed-point value parameter.

Since most astrophysical BHs are expected to be rotating~\cite{Bardeen:1970zz,Thorne:1974ve,Gammie:2003qi}, and X-ray spectroscopy is at its best for rapidly rotating BHs~\cite{Dauser:2013xv}, we shall focus on rotating BH solutions in asymptotically safe gravity. The rest of the paper is organized as follows: in Sec.~\ref{s-metric}, we review the BH solutions in asymptotically safe gravity. A review of X-ray spectroscopy, the \textsc{relxill\_nk} framework, and the numerical techniques used is provided in Sec.~\ref{s-xrs}. Sec.~\ref{s-analysis} details the X-ray source, observation and data analysis. Data analysis results are discussed and a comparison with previous works is presented in Sec.~\ref{s-discuss}.

\section{The metric \label{s-metric}}

To arrive at BH solutions in asymptotically safe gravity,~\cite{Held:2019xde} prescribes two steps. Firstly, the Newton's Coupling constant $G_N$ is replaced with the generalized length dependent $G_N(k)$, to include effects of quantum fluctuations of gravity. Secondly, the Renormalization Group energy scale $k$ is identified with a characteristic scale of the classical spacetime. This gives a modified BH metric. We follow this prescription now, as given in~\cite{Held:2019xde}, to generalize the Kerr metric. The usual Kerr metric in Boyer-Lindquist coordinates is given as
\bead\label{eq:metric}
	ds^2 = &-\frac{\Delta_r - a^2 \sin^2{\theta}}{\rho^2} dt^2 + \frac{\rho^2}{\Delta_r}dr^2 + \rho^2 d\theta^2 \\
	& + \frac{(a^2+r^2)^2 - a^2\Delta_r \sin^2\theta}{\rho^2}\sin^2\theta \,d\phi^2 \\
	& -\frac{2(a^2+r^2-\Delta_r)}{\rho^2}a \sin^2\theta \,dt\, d\phi
\eead
Here, $a$ is the specific angular momentum, defined as $a = J/M$, and
\begin{gather}
	\rho^2 = r^2 + a^2\cos^2\theta, \\
	\Delta_r = r^2 + a^2 - 2GMr.\label{eq:delta}
\end{gather}
The first step is to generalize $G$ to $G_N(k)$, which is given as
\be\label{eq:gn}
	G_N(k) = \frac{G_0}{1+\gamma G_0 k^2}. 
\ee
The second step is to identify $k$ with a scale in the spacetime, which is done as follows
\be\label{eq:ksq}
	k^2 = \frac{G_0Mr^3}{\rho^6}.
\ee
Thus, our modified-Kerr metric has one extra free parameter $\gamma$, the inverse dimensionless fixed-point value ($\gamma = g_{\star}^{-1}$).

We will work in the so called natural units. In these units, the gravitational radius is given as
\be\label{eq:rhor}
	r_g = M/M^2_{\rm Pl},
\ee
and
\be\label{eq:g0}
	G_0 = 1/M^2_{\rm Pl},
\ee
where $M$ is the BH mass parameter, and $M_{\rm Pl}$ is the Planck mass. Finally, we define a scaled $\gamma$ parameter as follows
\be\label{eq:tildegamma}
	\widetilde{\gamma} = \frac{\gamma M^{2}_{\rm Pl}}{M^2}.
\ee
For the rest of the paper, we will use $\widetilde{\gamma}$ as our non-Kerr deviation parameter.

\section{X-ray reflection spectroscopy}
\label{s-xrs}

\subsection{Theory}
\begin{figure}[!htb]
		\centering
		\includegraphics[width=0.48\textwidth]{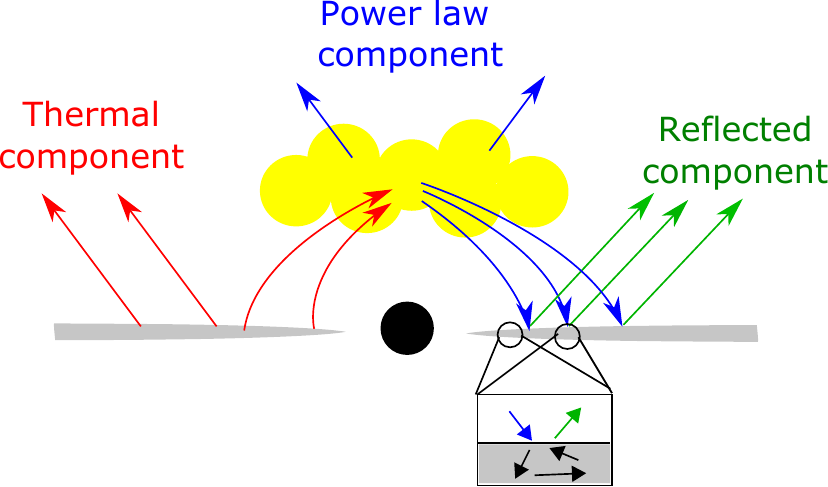}
		\caption{A schematic diagram of the disk-corona model. The central black circle denotes the BH. The disk is indicated in grey, and the corona in yellow. The coronal morphology is poorly understood so this illustration is only a guess. The arrows indicate photons and are colored according to the classification labeled on the figure and discussed in the text.}
		\label{fig:diskcorona}
\end{figure}
X-ray spectroscopy is based on the fundamental phenomenon of matter accreting on BHs. As matter falls from a companion star (in case of stellar-mass BHs) or galactic material (in case of supermassive BHs) into a BH, it heats up. During the accretion process, various high-energy processes lead to a plethora of radiation being emitted (See~\cite{Bambi:2018thh}, for instance, for a review.). These photons climb the gravitational well of the BH, traverse the Universe and are detected by X-ray telescopes stationed around the Earth. Effects of the interstellar/intergalactic media, the BH neighborhood and the BH itself are imprinted on the observed spectra. 

Fig.~\ref{fig:diskcorona} presents a schematic diagram of the BH neighborhood. The BH is at the center of the system, and in our case the spacetime metric is given by Eq.~\ref{eq:metric}. The disk is the standard Novikov-Thorne type disk~\cite{Novikov1973} with the following properties in particular: optically thick, geometrically thin, equatorial and no magnetic fields. (For all disk properties, see the reference above.) The size of disk is parametrized with $r_{\textrm{in}}$ and $r_{\textrm{out}}$ which quantify the inner and the outer edge of the disk, respectively. Additionally, the system includes a ``corona''. It is a source of very high energy photons, with effective temperatures of the order of 100~keV, compared to the disk where effective temperatures are given as~\cite{Bambi:2018thh}
\be\label{eq-mM}
T_{\rm eff} \sim \left ( \frac{10}{M} \right )^{1/4} {\rm keV},
\ee
where $M$ is the BH mass in the units of solar mass, and are of the order of $1$ keV for stellar-mass BHs and $0.01$ keV for supermassive BHs. The formation and morphology of the corona is a topic of active research and is probably different for different sources~\cite{Bambi:2018thh,Wilkins2012,Wilkins2014} (e.g., it could be the base of an astrophysical jet~\cite{Kara:2016kbu}, or a ring of high energy electrons above the accretion disk~\cite{Wilkins:2011kt}). 
 
Given the plethora of high-energy processes happening in the BH neighborhood, the observed spectra is mixture of radiation of varied origins. For the model described in Fig.~\ref{fig:diskcorona}, the total spectrum has three components. Firstly, the particles in the disk give off thermal radiation. As particle speeds vary along the disk, the total thermal radiation is given by a multicolor blackbody radiation~\cite{Bambi:2012tg}. Secondly, some of these thermal photons interact with the corona and, as a result of Compton upscattering, contribute a \textit{power-law} component to the total spectrum. Thirdly, those upscattered photons that impinge the disk, get reprocessed and are reflected back, giving rise to a \textit{reflected} component. Among the three, the power-law component is not very informative about the metric, and the thermal component is only weakly informative~\cite{Bambi2015}. The reflected component is most important for our purpose, and we describe now how the asymptotically free metric shown in Eq.~\ref{eq:metric} was implemented in a reflection model.

\subsection{The RELXILL\_NK model \label{s-trf0}}
The reflection component is sensitive to not just the BH (which determines particle motion in the disk and the photon trajectory from the point of emission to the telescope) but also to the structure and composition of the disk, as well as the corona. Reflection models therefore must include parameters related to all the aspects of the disk-corona model. To this end, a suite of models, called \textsc{relxill\_nk}~\cite{relxillnk,Abdikamalov:2019yrr,relxillnkweb,relxillnkweb2}, has been developed. It is built for the standard X-ray data analysis software XSPEC and includes a large class of BH-disk-corona models. It can model both the reflection and the power-law components of the spectrum. The eponymous model is described below.

Tab.~\ref{tab:params} lists the basic \textsc{relxill\_nk} model parameters and their default values. These parameters describe different aspects of the system, as follows. 
\bitem
	\item The spacetime is modeled using three parameters: $a_*$ specifies the BH spin, $\delta$-type is an integer that is used to switch between different non-Kerr deviation parameters, and $\delta$-value specifies the value of the chosen parameter.\footnote{Note that the BH mass is not a model parameter since, unlike the thermal spectrum, the reflection spectrum does not  depend on the BH mass explicitly.}
	\item The emissivity profile of the disk is modeled as a power law as follows:
\bead\label{eq:emis}
	I &\propto \frac{1}{r^{q_{\rm in}}} \qquad {\rm if} \quad r < r_{\rm br}, \nonumber \\
	I &\propto \frac{1}{r^{q_{\rm out}}} \qquad {\rm if} \quad r \geq r_{\rm br}. \nonumber
\eead
{where $r_{\rm br}$ is some radius of transition from one power-law index to another.} 
	\item {The following assumptions are made about the disk structure: it is assumed to be infinitesimally thin, confined in the equatorial plane and particles in the disk move in quasi-geodesic circular orbits. Thus only two structure parameters are needed to describe the disk structure, viz. $r_{\rm in}$ and $r_{\rm out}$, the inner and the outer radius of the disk, respectively. The inner bounds of the disk are taken to be at the innermost stable circular orbit (ISCO hereafter), since there are no stable circular orbits below this bound and in steady state very little radiation is expected to emerge from below this bound, and the outer bound is effectively set by the decaying emissivity.}
	\item The composition of the disk is assumed to follow our sun, i.e., the relative elemental abundances follow their solar values. The notable exception is iron, which is modeled with $A_{\rm Fe}$, defined as the ratio of iron content in the disk and the iron content in the sun. This is to account for higher (or lower) iron content in the accretion disks, as it depends on the history of the constituents that make up the disk. Besides this, $\log\xi$ parametrizes the ionization of the disk  (where $\xi$ is in units of erg\:cm/s), and ranges from 0 (neutral) to 4.7 (highly ionized).
	\item The coronal emission is taken care of with $\Gamma$, which is the index of the power-law component, and the high energy cut-off $E_{\rm cut}$, beyond which the power-law component is 
	exponentially suppressed. The latter is a feature that can be inferred from observations and must be included if we have data covering the hard X-ray spectrum. 
	\item Since the model includes both the power-law and the reflection component, $R_f$ is provided to control the relative contributions of the two components. It is defined as the ratio of intensity emitted towards the disk and that escaping to infinity.
	\item The observer's viewing angle is accounted with $i$ and the overall normalization with $N$.
\eitem 
Other models in the \textsc{relxill\_nk} suite change one or more aspect of the basic model, e.g., \textsc{relxilllp\_nk} assumes the corona is a point source on the BH spin axis, \textsc{relxillD\_nk} allows for higher electron density in the disk, and so on. For a complete list of models in the \textsc{relxill\_nk} suite, please see~\cite{Abdikamalov:2019yrr}.
\begin{table*}[!htb]
 \begin{center}
  \begin{tabular}{m{4cm} >{\centering\arraybackslash}m{2.5cm} >{\centering\arraybackslash}m{2.5cm} >{\centering\arraybackslash}m{2.5cm} >{\centering\arraybackslash}m{2.5cm}} \\
   \hline
   \textbf{Model} & Default & Left & Right & \textbf{Best-fit} \\
   \hline \hline
   \textsc{tbabs} \\
   $N_H/10^{22} cm^{-2}$ & -- & -- & -- & $8.84^{+0.06}_{-0.04}$  \\
   \hline
   \textsc{relxill\_nk} \\
   $q_{\rm in}$ &  $3$ &  $3$ &  $9$ &  $9.77_{-0.21}^{\rm + (P)}$ \\\vspace{0.2cm}
   $q_{\rm out}$ &  $3$ &  $3$ &  $3$ &  $0.0^{+0.3}$\\\vspace{0.2cm}
   $r_{\rm br}$ [$M$] & $15$ & $15$ & $15$ & $6.41^{+0.19}_{-0.60}$\\\vspace{0.2cm}
   $a_*$ &  $0.998$ &  $0.75$ &  $0.99$ &  $0.972^{+0.022}_{-0.019}$ \\\vspace{0.2cm}
   $i$ [deg] &  $30$ &  $30$ &  $30$ &  $73.8^{+0.7}_{-0.4}$ \\\vspace{0.2cm}
   $\Gamma$ & $2$ & $2$ & $2$ & $2.39^{+0.05}_{-0.03}$ \\\vspace{0.2cm}
   $\log\xi$ &  $3.1$ &  $3.1$ &  $3.1$ &  $2.72^{+0.03}_{-0.02}$ \\\vspace{0.2cm}
   $A_{\textrm{Fe}}$ &  $1$ &  $5$ &  $5$ &  $0.53^{+0.06}_{\rm - (P)}$ \\\vspace{0.2cm}
   $E_{\rm cut}$ [keV] & $300$ & $300$ & $300$ & $122^{+13}_{-6}$\\\vspace{0.2cm}
   $R_f$ &  $3$ &  $3$ &  $3$ &  $0.79^{+0.05}_{-0.04}$ \\\vspace{0.2cm}
   $\delta$-type &  $\alpha_{13}$ &  $\widetilde{\gamma}$ &  $\widetilde{\gamma}$  &  $\widetilde{\gamma}$  \\\vspace{0.2cm}
   $\delta$-value &  $0$ &  $0/0.1/0.2$ &  $0/0.0035/0.007$ &  $0.024^{+ 0.023}_{\rm - (P)}$ \\
     \hline\vspace{0.2cm}
   $\chi^2/dof$  & -- & -- & -- & $2301.03/2222$ \\
   & & & & $= 1.03557$ \\
   \hline
   \hline
  \end{tabular}
    \caption{{Summary of the model parameters and their values for different configurations. The second column lists the default values in the model and is discussed in Sec.~\ref{s-xrs}. The third and fourth columns list their values for the qualitative discussion in Sec.~\ref{s-qualitative}. The last column lists the best-fit values for the best-fit spectral model discussed in Sec.~\ref{s-analysis}. The reported uncertainties correspond to the 90\% confidence level for one relevant parameter. For $q_{\rm out}$ there is no lower uncertainty because the best-fit value is stuck at the lower boundary. (P) indicates that the 90\% confidence level reaches one of the boundaries of the parameter: the upper boundary of $q_{\rm in}$ is 10, the lower boundary of $A_{\textrm{Fe}}$ is 0.5, and the lower boundary of $\widetilde{\gamma}$ is 0. Apart from those listed, the model has the following extra parameters which are set to their default value in all cases: $r_{\rm in}$ ($=$ ISCO), $r_{\rm out}$ ($=400$).}}
    \label{tab:params}
 \end{center}
\end{table*}

\subsection{Numerical method \label{s-trf}}
We now describe how the metric in Eq.~\ref{eq:metric} was implemented in the \textsc{relxill\_nk} model. {The methodology described here is standard~\cite{Fabian:1989ej,Laor:1991nc,speith1995} and we follow the terminology of~\cite{relxillnk} and~\cite{Abdikamalov:2019yrr}. The flux received on a telescope screen can be written as}
\be\label{eq-Fobs1}
F_o (\nu_o) 
= \int I_o(\nu_o, X, Y) d\tilde{\Omega}\, ,
\ee
where $I_{ o}$ is the specific intensity (e.g., in units of erg~s$^{-1}$~cm$^{-2}$~str$^{-1}$~Hz$^{-1}$) at the telescope screen, $X$ and $Y$ are Cartesian coordinates on the telescope screen, and $d\tilde{\Omega} = dX dY/D^2$ is the solid angle element subtended by the disk on the telescope sky. $d\tilde{\Omega}$ can be rewritten in terms of the redshift factor $g$ and the transfer function $f$. The redshift factor $g$ is defined as 
\be
	g = \frac{\nu_o}{\nu_e}.
\ee
where $\nu_o$ is the photon's frequency in the telescope's frame of reference at the telescope, and $\nu_e$ the photon's frequency in the emitter's rest frame at the disk.
The transfer function, first introduced in Ref.~\cite{Cunningham1975}, is defined as: 
\be\label{eq-trf}
f(g^*,r_e,i) = \frac{1}{\pi r_e} g 
\sqrt{g^* (1 - g^*)} \left| \frac{\partial \left(X,Y\right)}{\partial \left(g^*,r_e\right)} \right| \, .
\ee
where $r_e$ is the radial coordinate on the disk and $g^*$ is a normalized redshift factor. It is defined as
\be
g^* = \frac{g - g_{\rm min}}{g_{\rm max} - g_{\rm min}} \, ,
\ee
where $g_{\rm max}=g_{\rm max}(r_e,i)$ and $g_{\rm min}=g_{\rm min}(r_e,i)$ are, respectively, the maximum and the minimum values of the redshift factor $g$ at a fixed $r_e$ and for a fixed inclination angle of the telescope relative to the BH spin axis, and it divides a constant $r_e$ ring on the disk in two branches, bounded by $g^*=0$ and $g^*=1$.
Furthermore, $I_{ o}$ can be recast using the Liouville's theorem: 
\be
I_o = g^3 I_e,
\ee
in terms of the redshift factor and $I_e$, the specific intensity at the point of emission. This results in the following expression for the flux. 
\bwide
\be\label{eq-Fobs2}
F_o (\nu_o) 
= \frac{1}{D^2} \int_{r_{\rm in}}^{r_{\rm out}} \int_0^1
\pi r_{ e} \frac{ g^2}{\sqrt{g^* (1 - g^*)}} f(g^*,r_{e},i)
I_{  e}(\nu_{ e},r_{ e},\vartheta_{ e}) \, dg^* \, dr_{ e} \, .
\ee
\ewide
Here $D$ is the distance of the source from the observer and $\vartheta_{e}$ is the photon's direction relative to the disk when it is emitted. $r_{\rm in}$ and $r_{\rm out}$ are the inner and the outer edges of the accretion disk, respectively.
 
The introduction of the transfer function enables a separation between the microphysics at the disk and the photon travel along the null geodesic. The reflection spectrum can be readily calculated using Eq.~\ref{eq-Fobs2} if the transfer function is known. But the high computational cost to calculate the transfer function by tracing photons and using Eq.~\ref{eq-trf} whenever the flux needs to be calculated, prohibits the direct usage these equations. 
Rather, the \textsc{relxill\_nk} framework uses interpolation schemes to calculate the transfer function for any $\{g^*,r_e,i\}$ from the transfer functions for some $\{g^*,r_e,i\}$. The transfer function data for some $\{g^*,r_e,i\}$ is stored in a FITS (Flexible Image Transport System) table. The procedure to create such a table is described in~\cite{relxillnk,Abdikamalov:2019yrr}. We briefly overview this scheme here. 

\begin{figure}[!htb]
		\centering
		\includegraphics[width=0.5\textwidth]{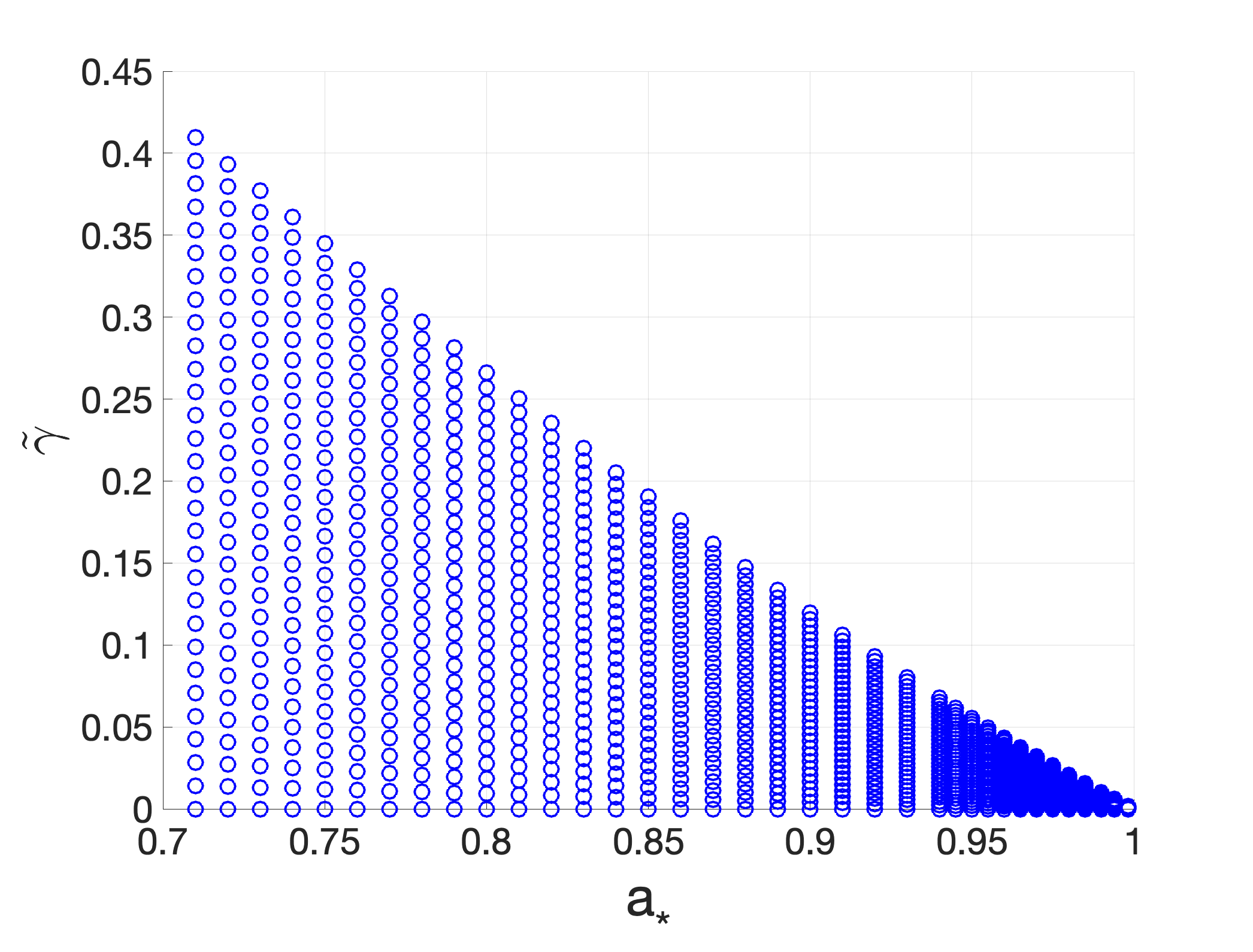}
		\caption{The grid of values, represented by blue circles, of {$a_*$ and $\widetilde{\gamma}$} for which the transfer functions are calculated and stored in the FITS table. Note that the grid spacings are non-uniform in both $a_*$ and $\widetilde{\gamma}$.}
		\label{fig:spindefpargrid}
\end{figure}

The three model parameters spin $a_*$, {non-Kerr deviation} parameter $\widetilde{\gamma}$ and the telescope's inclination angle $i$, are discretized in a $36\times30\times22$ grid, respectively. The grid spacing in each dimension is non-uniform, e.g., the grid becomes denser as $a_*$ increases, since the ISCO radius changes faster with increasing $a_*$. The spacing is chosen as a balance between sufficient resolution during interpolation and a reasonable FITS file size. The range of $a_*$ is from $0.71$ to $0.9982$, since our focus is on rapidly rotating BHs. The range of $\widetilde{\gamma}$ is obtained by imposing that our metric describes a black hole with an event horizon, namely the equation $\Delta_r = 0$ has at least a real positive solution. Such a requirement gives an upper bound on $\widetilde{\gamma}$, say $\widetilde{\gamma} \le \widetilde{\gamma}_{\rm crit}$. The lower bound is simply zero, following Eq.~\ref{eq:tildegamma} and since all quantities therein are greater than or equal to zero. The Kerr solution is recovered at $\widetilde{\gamma}=0$. The $36 \times 30$ grid for $a_* $ and $\widetilde{\gamma}$ is shown in Fig.~\ref{fig:spindefpargrid}.

For each grid point (i.e., for each $a_*$, $\widetilde{\gamma}$, and $i$), the accretion disk is discretized in the $r_e$ and the $g^*$ dimension, with 100 and 40 values, respectively.\footnote{Because of the way the transfer function is defined in Eq.~\ref{eq-trf}, it goes to zero when the redshift is maximum or minimum, resulting in two branches of transfer function between $g^*=0$ and  $g^*=1$.} The $r_e$ dimension ranges from the ISCO to $1000M$, and is non-uniform, with higher density near the ISCO. The $g^*$ dimension is equally spaced between $\epsilon$ and $1-\epsilon$, where $\epsilon = 10^{-3}$ (and not $0$ and $1$ since the transfer function diverges at those values). A fourth order Runge-Kutta ray-tracing scheme, described in~\cite{relxillnk,Abdikamalov:2019yrr}, traces the photons backwards in time from the telescope screen (placed at asymptotically large distance from the BH) to the disk. Due to the highly curved spacetime near the BH, the landing location of the photon on the disk is not known a priori. An adaptive algorithm is used to fine-tune the initial position of the photon on the telescope screen so that the photon, when ray-traced backwards, lands at the desired $r_{\textrm{e}}$. For each such ``central'' photon, the code calculates the redshift, emission angle, etc. needed for calculation of the transfer function using Eq.~\ref{eq-trf}. Four additional photons are fired for each central photon to calculate the Jacobian in Eq.~\ref{eq-trf}. The initial positions of the additional photons on the telescope screen are chosen to ensure that the resultant Jacobian is convergent. For each $r_{\textrm{e}}$, about 100 central photons are computed, which are then interpolated to get the transfer function and the emission angle on the 40 equally spaced values of $g^*$, which is stored in the FITS file.

{\section{Effect of $\widetilde{\gamma}$ on the observables\label{s-qualitative}}

Before delving into the quantitative analysis of the fixed point parameter $\widetilde{\gamma}$, we look at its effects on some observables in this section, in particular we identify scenarios which will provide a good measurement of $\widetilde{\gamma}$.} 

\begin{figure}[!htb]
		\centering
		\includegraphics[width=0.5\textwidth]{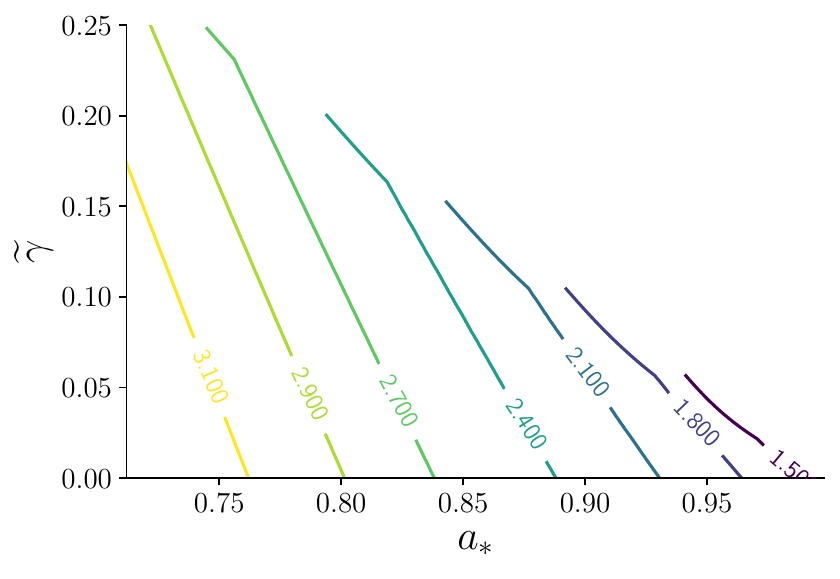}
		\caption{{Constant ISCO contours in the spin-$\widetilde{\gamma}$ phase space under consideration. The contours are labeled by the ISCO radius along them. The top right half region of the phase space is excluded as it violates the $\widetilde{\gamma} \le \widetilde{\gamma}_{\rm crit}$ condition (see Sec.~\ref{s-trf}).}}
		\label{fig:iscocontours}
\end{figure}

{One of the most important quantities in the context of accretion disks and X-ray spectroscopy is the ISCO. It is closely related to the properties of the BH, forms the inner edge of the disk in our model\footnote{While the assumption of the inner edge of the disk being located at the ISCO is well-motivated theoretically~\cite{Reynolds:2007rx,Penna:2010hu} and observationally~\cite{Fabian:1989ej,Steiner:2010kd}, in some cases the disk may truncate at a larger radius~\cite{Done:2007nc,Zdziarski:2020pgi} and result in a systematic error on parameter estimates. Overcoming this systematic uncertainty may become possible in future~\cite{Wilkins:2020pgu}.} and so can be directly measured, and has a very strong effect on the radiation profile. For the region of the $(a_*, \widetilde{\gamma})$ phase space which we are considering in this work (See Fig.~\ref{fig:spindefpargrid}), the ISCO contours are plotted in Fig.~\ref{fig:iscocontours}. Generically, we find that increasing $\widetilde{\gamma}$ at a constant spin reduces the ISCO radius. This effect is stronger at higher spins. Thus, we can anticipate faster spinning BHs to provide stronger constraints on $\widetilde{\gamma}$. No such effect on $\widetilde{\gamma}$ is expected due to the BH mass, since the BH mass plays no role in X-ray reflection spectroscopy. But since the physical fixed-point value parameter $\gamma$ is directly proportional to the square of the BH mass (See Eq.~\ref{eq:tildegamma}), it will be constrained better for lower-mass BHs. }
Another feature we can observe in Fig.~\ref{fig:iscocontours} is the dependency of the ISCO radius on both $\widetilde{\gamma}$ and the BH spin. This translates to a partial degeneracy between the two parameters and weakens any constraint on either parameters. For instance, a high-spin pure Kerr measurement can be degenerate with a lower-spin non-Kerr measurement, as we will see in Sec.~\ref{s-discuss} when we analyze astrophysical data.
 \begin{figure*}[!htb]
		\centering
		\includegraphics[width=\columnwidth]{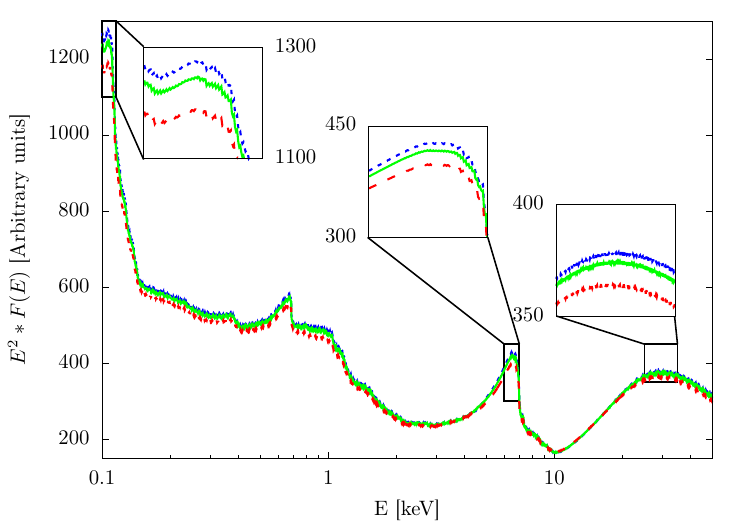}
		\includegraphics[width=\columnwidth]{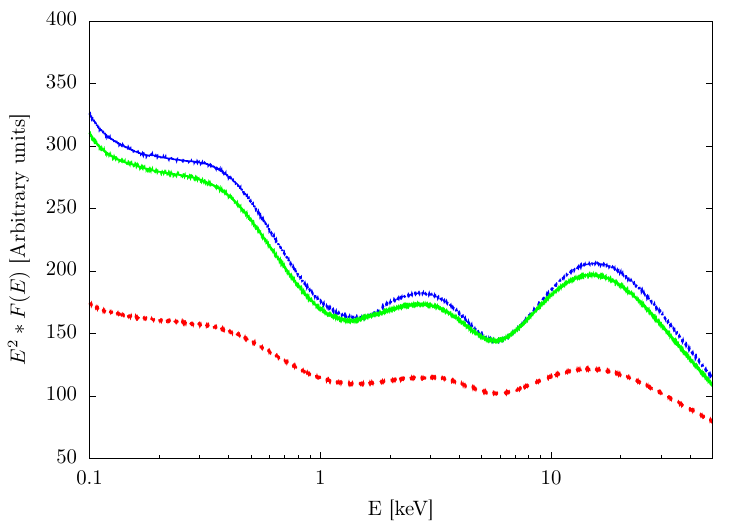}
		\caption{{Illustrative reflection spectra, obtained from \textsc{relxill\_nk}, for two values of BH spin and emissivity index, and various values of $\widetilde{\gamma}$. Left: $a=0.75$, $q_{\rm in}=3$, $\widetilde{\gamma}=0$ (blue), $0.1$ (green), $0.2$ (red). Right: $a=0.99$, $q_{\rm in}=9$, $\widetilde{\gamma}=0$ (blue), $0.0035$ (green), $0.007$ (red). Insets in the left panel zoom in on the soft-excess, K$\alpha$ iron and Compton hump sectors, respectively, of the reflection spectrum. Other model parameters are set to default (see Tab.~\ref{tab:params}). See the text for more details.}}
		\label{fig:refspec}
\end{figure*}

{The next observable we look at is the reflection spectrum. This is the output of the \textsc{relxill\_nk} model and gives us a clear idea of the measurability of $\widetilde{\gamma}$ with the X-ray reflection technique. Fig.~\ref{fig:refspec} shows the flux for representative BH-disk configurations. Parameter values for each case are given in the third and fourth columns in Tab.~\ref{tab:params}. In particular, in the panel on left, $a_*=0.75$, which is close to the lowest spin under consideration. In the same panel, $\widetilde{\gamma} \simeq (0, 0.1, 0.2)$ in the three curves, respectively, which span the complete range of $\widetilde{\gamma}$ at that spin (see Fig~\ref{fig:spindefpargrid}). The iron abundance is set to $5$ to enhance the features associated with the iron emission profile\footnote{{Observationally too, it is common to find super-solar abundances in astrophysical sources (see, e.g., Fig. 3 in~\cite{Garcia:2018czl}). Whether such iron fractions are real, and not a result of model shortcomings~\cite{Reynolds:2012ji,Garcia:2018czl}, is a matter of active research.}}, and rest of the parameters are set to their default values. The primary features of a typical reflection spectrum, e.g., a broad iron line around $6-7$ keV, a Compton hump at $20-50$ keV, and a high photon count below $1$ keV, are all present. The most important feature though is the minute effect of the $\widetilde{\gamma}$ parameter on the whole spectrum. Only below $0.1$ keV, the effect is somewhat noticeable. Even a $15\%$ change in the ISCO radius between $\widetilde{\gamma} \simeq 0$ and $0.2$, and a high iron abundance, does not make a big difference. At higher spins, though the ISCO is smaller and more sensitive to $\widetilde{\gamma}$, the allowed range of $\widetilde{\gamma}$ is small, and the reflection spectrum continues to be nearly unchanged for the whole range of allowed $\widetilde{\gamma}$. Overall, present X-ray instruments are unlikely to provide data that can distinguish this minute effect of the $\widetilde{\gamma}$ parameter. Even with future instruments, only in ideal conditions (bright and steady source, high BH spin, high resolution in the very low energy band, etc.) is there a possibility of constraining $\widetilde{\gamma}$.}

{It is well known that the reflection spectra is quite sensitive to the emissivity profile. While making the observations about the effect of $\widetilde{\gamma}$ on the reflection spectra above, we chose the default emissivity profile of the model. But it is common in astrophysical systems to have a steeper emissivity profile~\cite{Fabian:2002gj,Waddell:2020dqr}. With this in mind, in the panel on right in Fig.~\ref{fig:refspec}, we plot the flux with a steep emissivity profile. We set the spin to a high value, and span the whole range of $\widetilde{\gamma}$ at that spin ($\widetilde{\gamma} \simeq 0, 0.0035, 0.007$). Rest of the parameters are the same as in the left panel and listed in the fourth column in Tab.~\ref{tab:params}. We now see a clear change in the reflection spectrum with $\widetilde{\gamma}$. The effect is particularly strong at lower energies, in the so called soft-excess, and} gets stronger as the fixed-point value moves further away from zero. Thus, we anticipate sources with steep emissivity profile to be suitable for testing asymptotically safe quantum gravity.

\section{Data analysis\label{s-analysis}}

Having developed the model to calculate reflection spectra for the BH in Eq.~\ref{eq:metric}, we used this model to analyze data from an X-ray binary to constrain the free parameter $\widetilde{\gamma}$.
In this section, we describe the X-ray source, the observation we used to analyze this source, and the data analysis.

{The data that we are going to analyze are from a {\em Suzaku} observation of GRS~1915+105 in 2007. While GRS~1915+105 is normally a variable and quite complicated source, in that 2007 observation it was stable, presented a simple spectrum with strong reflection features, the Eddington-scaled disk luminosity was around 20\% and this guarantees a thin accretion disk, and the temperature of the disk was low, see~\cite{Zhang:2019ldz}. The BH in GRS~1915+105 has a very high spin and the inclination angle of the accretion disk with respect to the line of sight of the distant observer is high which, combined with the fact that the inner edge of the disk was well illuminated during the 2007 {\em Suzaku} observation, maximize the relativistic effects in the reflection spectrum. Last, {\em Suzaku} permits to have both a good energy resolution near the iron line and data up to 50~keV to observe the Compton hump. All these ingredients make this particular observation of GRS~1915+105 quite suitable for GR tests.}


\subsection{Review \label{ss:ana-rev}}
GRS 1915+105 (also known as V1487 Aquilae) is a low mass X-binary at a distance of 8.6 kiloparsecs~\cite{Reid:2014ywa}. The mass of the BH in GRS 1915+105 is $\sim12.4 M_{\odot}$, making it one of the most massive stellar BHs detected in the Milky Way galaxy. It has been a persistent source of X-rays since 1992. It has been extensively analyzed with the \textsc{relxill\_nk} suite of models, with two observations, one from the {\em NuSTAR} telescope and the other from the {\em Suzaku} telescope. 
In~\cite{Zhang:2019zsn}, we used \textsc{relxill\_nk} to analyze the 2012 {\em NuSTAR} observation. In~\cite{Zhang:2019ldz}, we analyzed the 2007 {\em Suzaku} observation. 
The former observation proved difficult to fit and gave inconsistent values of the {non-Kerr deviation} parameters analyzed there. The {\em Suzaku} observation on the other hand required fewer models for a good fit, and gave consistent best-fit values for the parameters. A difference in the state of the source was also inferred from the observations. The {\em NuSTAR} observation required a thermal component, suggesting a hotter disk during the observation. The {\em Suzaku} observation, on the other hand, required no thermal component, suggesting a colder disk during the observation. Since the \textsc{relxill\_nk} model, which based on \textsc{xillver}, assumes a cold disk, the fits with the {\em Suzaku} observation were deemed more trustworthy. 

In~\cite{Zhang:2019ldz}, the {\em Suzaku} observation was also analyzed with different versions of \textsc{relxill\_nk} and different {non-Kerr deviation} parameters. A qualitative picture emerged thus: the base \textsc{relxill\_nk} model provides a good fit to the observation, the emissivity profile follows the broken power law, shown in Eq.~\ref{eq:emis}. The emissivity index in the inner parts of the disk is quite large and in the outer parts it is quite small (Such an emissivity profile could be a result of a ring-like corona above the accretion disk~\cite{Miniutti:2003yd,Wilkins:2011kt}), the BH spin is very high ($\sim 0.99$), and the inclination is $\sim 60-70$ degrees. Different {non-Kerr deviation} parameters have been tested with this observation~\cite{Zhu:2020cfn,Zhang:2019ldz}. In all cases, the Kerr limit was covered with high confidence, making the BH consistent with a Kerr BH. Recently, this observation was analyzed with a thick-disk version of \textsc{relxill\_nk}~\cite{Abdikamalov:2020oci} (typical models assume an infinitesimally thin disk, this assumption was relaxed in this study). The thick disk \textsc{relxill\_nk} model provides only a marginally better fit than the base \textsc{relxill\_nk} model, with no significant difference in the best-fit values of the model parameters. {This observation was also used by some of us to study an important systematic effect due to the assumption that particles in the disk move on nearly-geodesic circular orbits~\cite{Tripathi:2020wfi}, where we found this assumption can be put to test in the presence of a robust independent estimate of the disk's inclination relative to the observer.}

\subsection{Observations and data reduction \label{s-red}}
The {\em Suzaku} observation we will analyze was made on May 7, 2007 for 117~kiloseconds (Obs ID 402071010). Among the four XIS units on board, two were turned off to preserve telemetry and a third unit was running in the timing mode. Therefore we have data from XIS1 only. Additionally, we have data from the HXD/PIN instrument on board. XIS0 and XIS2 are turned off during this period and XIS3 has yet to be calibrated.

The details of data reduction for this observation have been given in~\cite{Zhang:2019ldz}. We only mention some highlights of the reduced data here, and refer the reader to that reference for further details. The data was grouped to have a minimum of 25 photons per bin. For XIS1, a net exposure time of 28.94~ks in the $3\times 3$ editing mode was achieved. The energy band for data analysis was finalized as $2.3-10$ keV. The lower bound is due to a lack of sufficient photons below $2.3$ keV, and the upper bound is chosen in order to avoid calibration issues near the Si~K edge. For the HXD/PIN data, a net exposure time of 53.00~ks was achieved in the 12.0--55.0 keV energy band (the choice of the energy range follows~\cite{Blum:2009ez}).
         XSPEC 12.10.1f was used to analyze the data.

\subsection{Modelling and results \label{s-red}}
Since this observation has been analyzed before with {\sc relxill\_nk}, it was natural to begin with the best-fit model combination found previously as a first guess. We thus fit the observation with the following model:

\vspace{0.07in}
{\centering
\sc {Model: constant*tbabs*relxill\_nk}.\par}
\vspace{0.05in}

\noindent A constant is allowed to float between two data sets. It is frozen to 1 for XIS1 and thawed for HXD to account for the cross-calibration constant between XIS1 and HXD. Here \textsc{tbabs} accounts for galactic absorption~\cite{Wilms2000}. The galactic column density parameter is kept free during the fitting. The power-law and the reflection components are modeled with \textsc{relxill\_nk}. The inner edge of the disk is assumed to be at the innermost stable circular orbit, a standard assumption valid in particular for this observation since the Eddington scaled accretion luminosity was ~20\% during the observation~\cite{Blum:2009ez,Steiner:2010kd,Kulkarni:2011cy}, and the outer edge at $400M$ (the fit is not particularly sensitive to the outer radius and therefore we leave it at its default value). 

 \begin{figure}[!htb]
		\centering
		\includegraphics[width=0.48\textwidth]{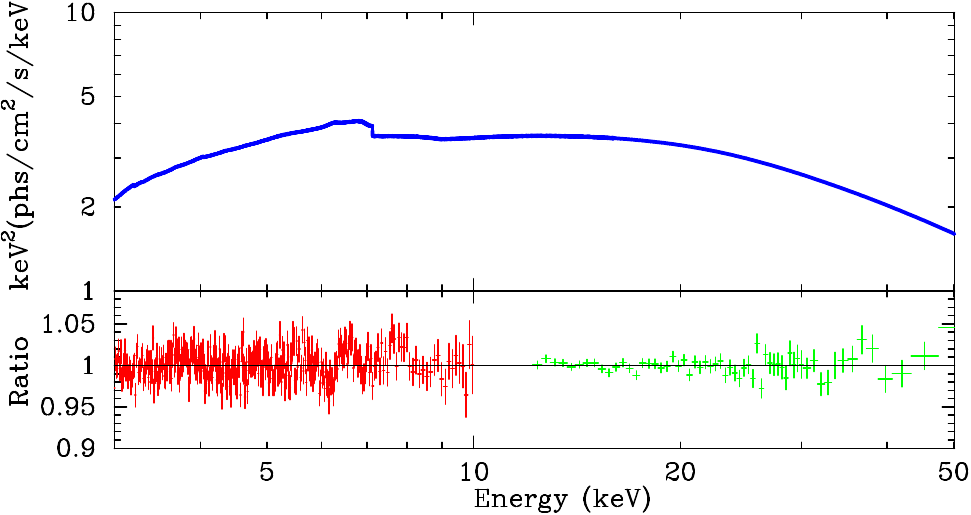}
		\caption{Top: The spectral model presented in Tab.~\ref{tab:params}. Bottom: Data to model ratio for the spectral fit shown in the top panel. The XIS1 data are in red, the HXD/PIN data in green. See the text for more details.}
		\label{fig:ratio}
\end{figure}
The best-fit model and the ratio of the data to the model is shown in Fig.~\ref{fig:ratio}, in the top and the bottom panels respectively. Note that there are no significant features in the ratio plot, suggesting that the model fits the data well. Tab.~\ref{tab:params} shows the best-fit parameter values, {estimated for each parameter by marginalizing over the rest of the parameters,} and the $\chi^2$ measure of the best-fit. Since the reduced $\chi^2$ is close to 1, the best-fit model provides a good statistical fit to data. We can thus conclude that the model provides a satisfactory fit to the data.

\section{Discussion\label{s-discuss}}
We now discuss the results of the data analysis presented above. We can compare the best-fit parameter values obtained here with their values in other analyses. Among the BH neighborhood parameters, the emissivity profile, for example, follows previous results, with $q_{\rm in}$ pegged at a large value, and a nearly zero value of $q_{\rm out}$, with the break occurring near $6M$. Such an emissivity profile is expected for a ring-like corona above the accretion disk~\cite{Miniutti:2003yd,Wilkins:2011kt}. The spin and inclination are high, the iron abundance is below solar, and the cut off energy is quite low, all of which is consistent with previous results~\cite{Blum:2009ez,Miller:2013rca}.
  
Of course, the parameter of primary interest in this analysis is the {fixed-point value parameter $\widetilde{\gamma}$ (see Eq.~\ref{eq:tildegamma})}. We find that
\be\label{eq:cons}
	 \widetilde{\gamma} \slsim 0.047,
\ee
at $90\%$ confidence. This can be directly compared with the results in~\cite{Held:2019xde}. There, the authors, using bounds on the mass of M$87^*$ obtained by the EHT collaboration~\cite{Akiyama:2019eap}, provide the following constraint: $\widetilde{\gamma}\slsim 2$, which is significantly weaker than the constraints obtained in the present work. Future observations, of Sgr A$^*$, by EHT are expected to improve the constraints to be: $\widetilde{\gamma} \slsim 0.5$, which is still an order of magnitude weaker than those in Eq.~\ref{eq:cons}. 
The difference between our constraint and those from black hole imaging is even more remarkable in terms of $\gamma$ which is the real physical parameter for the theory. {As discussed in Sec.~\ref{s-qualitative}, lower-mass BHs provide better constraints on $\gamma$. Since} GRS~1915+105 is a stellar-mass black hole while Sgr A$^*$ and M$87^*$ are supermassive black holes of, respectively, some million and some billion Solar masses, {we find}
\be
\gamma = \frac{\widetilde{\gamma} M^2}{M_{Pl}^2} \slsim 6 \cdot 10^{76} \, .
\ee 
{which is much stronger than present ($2 \cdot 10^{95}$) and even future ($10^{89}$) constraints possible with BH imaging.}

\begin{figure}[!htb]
		\centering
		\includegraphics[width=0.48\textwidth]{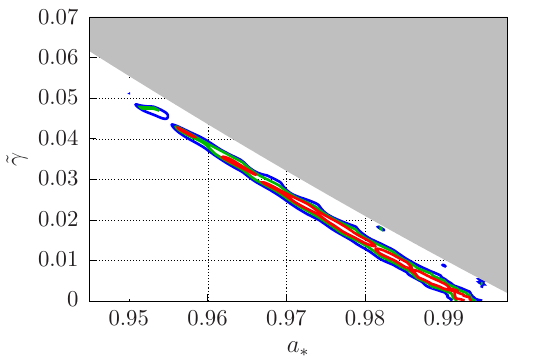}
		\caption{The contour plot of spin $a_*$ vs. $\widetilde{\gamma}$, illustrating the degeneracy between the two parameters. The red, green and blue lines show the 68\%, 90\%, and 99\% confidence level boundaries respectively. The Kerr solution is recovered at $\widetilde{\gamma}=0$. 
		The gray region is ignored in our analysis because in such a region the spacetime has no horizon ($\widetilde{\gamma} > \widetilde{\gamma}_{\rm crit}$, see Section~\ref{s-trf}).
		See the text for more details.}
		\label{fig:spindefpar}
\end{figure}

The results presented above should be seen in the proper context. {An important source of uncertainty in our measurement of $\widetilde{\gamma}$ is systematic error}. The \textsc{relxill\_nk} model makes a series of assumptions about the disk and the corona. Whether these assumption are valid depends on the particular source and the particular observation~\cite{Penna:2010hu,Zhang:2019ldz,Dauser:2013xv}. Even within the model, inter-parameter degeneracies may result in larger uncertainties. To illustrate this, let us take the example of the $\widetilde{\gamma}$ parameter. Following the discussion in Sec.~\ref{s-qualitative}, we can anticipate that for a disk with a steep emissivity profile, $\widetilde{\gamma}$ would be constrained very well at a given spin. For instance, if we assume the spin of GRS 1915+105 is given \textit{exactly} by 0.972 (See Tab.~\ref{tab:params}), $\widetilde{\gamma}\lesssim 0.02$ since that is that maximum range of $\widetilde{\gamma}$ at that spin (see Sec.~\ref{s-trf} and Fig.~\ref{fig:spindefpargrid}). But in reality, the BH spin and $\widetilde{\gamma}$ are degenerate to some extent, resulting in a weaker constraint on $\widetilde{\gamma}$. Degeneracy with other model parameters, e.g., the disk inclination relative to the observer and the emissivity index, further weakens the constraint on $\widetilde{\gamma}$. Marginalizing over all free model parameters is therefore critical to get reliable estimates, and the constraint reported in Eq.~\ref{eq:cons} properly accounts for this. To understand the driving factor behind the spin-$\widetilde{\gamma}$ degeneracy, we look at contours of $\chi^2$ for spin vs $\widetilde{\gamma}$ while marginalizing over all the other free parameters. The $68\%$, $90\%$ and $99\%$ confidence level contours for this degeneracy are shown in Fig.~\ref{fig:spindefpar}. Indeed, as discussed qualitatively in Sec.~\ref{s-qualitative}, a steep emissivity profile allows meaningful constraints on $\widetilde{\gamma}$ (see Fig.~\ref{fig:refspec}), but at the expense of the spectrum being dominated by radiation from the innermost part of the disk, which leads to a degeneracy between $\widetilde{\gamma}$ and spin that closely follows the ISCO contours (see Fig.~\ref{fig:iscocontours}).

\noindent {\bf Acknowledgments --}
{We thank the referee for his/her constructive comments which helped improve the manuscript.} This work was supported by the Innovation Program of the Shanghai Municipal Education Commission, Grant No.~2019-01-07-00-07-E00035, and the National Natural Science Foundation of China (NSFC), Grant No.~11973019. A.B.A. acknowledges the support from the Shanghai Government Scholarship (SGS). S.N. acknowledges support from the Alexander von Humboldt Foundation.


\bibliography{references}

\begin{thebibliography}{95}%
\makeatletter
\providecommand \@ifxundefined [1]{%
 \@ifx{#1\undefined}
}%
\providecommand \@ifnum [1]{%
 \ifnum #1\expandafter \@firstoftwo
 \else \expandafter \@secondoftwo
 \fi
}%
\providecommand \@ifx [1]{%
 \ifx #1\expandafter \@firstoftwo
 \else \expandafter \@secondoftwo
 \fi
}%
\providecommand \natexlab [1]{#1}%
\providecommand \enquote  [1]{``#1''}%
\providecommand \bibnamefont  [1]{#1}%
\providecommand \bibfnamefont [1]{#1}%
\providecommand \citenamefont [1]{#1}%
\providecommand \href@noop [0]{\@secondoftwo}%
\providecommand \href [0]{\begingroup \@sanitize@url \@href}%
\providecommand \@href[1]{\@@startlink{#1}\@@href}%
\providecommand \@@href[1]{\endgroup#1\@@endlink}%
\providecommand \@sanitize@url [0]{\catcode `\\12\catcode `\$12\catcode
  `\&12\catcode `\#12\catcode `\^12\catcode `\_12\catcode `\%12\relax}%
\providecommand \@@startlink[1]{}%
\providecommand \@@endlink[0]{}%
\providecommand \url  [0]{\begingroup\@sanitize@url \@url }%
\providecommand \@url [1]{\endgroup\@href {#1}{\urlprefix }}%
\providecommand \urlprefix  [0]{URL }%
\providecommand \Eprint [0]{\href }%
\providecommand \doibase [0]{http://dx.doi.org/}%
\providecommand \selectlanguage [0]{\@gobble}%
\providecommand \bibinfo  [0]{\@secondoftwo}%
\providecommand \bibfield  [0]{\@secondoftwo}%
\providecommand \translation [1]{[#1]}%
\providecommand \BibitemOpen [0]{}%
\providecommand \bibitemStop [0]{}%
\providecommand \bibitemNoStop [0]{.\EOS\space}%
\providecommand \EOS [0]{\spacefactor3000\relax}%
\providecommand \BibitemShut  [1]{\csname bibitem#1\endcsname}%
\let\auto@bib@innerbib\@empty
\bibitem [{\citenamefont {Will}(2014)}]{will2014}%
  \BibitemOpen
  \bibfield  {author} {\bibinfo {author} {\bibfnamefont {C.~M.}\ \bibnamefont
  {Will}},\ }\href {\doibase 10.12942/lrr-2014-4} {\bibfield  {journal}
  {\bibinfo  {journal} {Living Rev. Rel.}\ }\textbf {\bibinfo {volume} {17}},\
  \bibinfo {pages} {4} (\bibinfo {year} {2014})},\ \Eprint
  {http://arxiv.org/abs/1403.7377} {arXiv:1403.7377 [gr-qc]} \BibitemShut
  {NoStop}%
\bibitem [{\citenamefont {Bambi}(2017)}]{Bambi2015}%
  \BibitemOpen
  \bibfield  {author} {\bibinfo {author} {\bibfnamefont {C.}~\bibnamefont
  {Bambi}},\ }\href {\doibase 10.1103/RevModPhys.89.025001} {\bibfield
  {journal} {\bibinfo  {journal} {Rev. Mod. Phys.}\ }\textbf {\bibinfo {volume}
  {89}},\ \bibinfo {pages} {025001} (\bibinfo {year} {2017})},\ \Eprint
  {http://arxiv.org/abs/1509.03884} {arXiv:1509.03884 [gr-qc]} \BibitemShut
  {NoStop}%
\bibitem [{\citenamefont {Abbott}\ \emph {et~al.}(2019)\citenamefont {Abbott}
  \emph {et~al.}}]{LIGOScientific:2019fpa}%
  \BibitemOpen
  \bibfield  {author} {\bibinfo {author} {\bibfnamefont {B.}~\bibnamefont
  {Abbott}} \emph {et~al.} (\bibinfo {collaboration} {LIGO Scientific,
  Virgo}),\ }\href {\doibase 10.1103/PhysRevD.100.104036} {\bibfield  {journal}
  {\bibinfo  {journal} {Phys. Rev. D}\ }\textbf {\bibinfo {volume} {100}},\
  \bibinfo {pages} {104036} (\bibinfo {year} {2019})},\ \Eprint
  {http://arxiv.org/abs/1903.04467} {arXiv:1903.04467 [gr-qc]} \BibitemShut
  {NoStop}%
\bibitem [{\citenamefont {Carson}\ and\ \citenamefont
  {Yagi}(2020)}]{Carson:2019rda}%
  \BibitemOpen
  \bibfield  {author} {\bibinfo {author} {\bibfnamefont {Z.}~\bibnamefont
  {Carson}}\ and\ \bibinfo {author} {\bibfnamefont {K.}~\bibnamefont {Yagi}},\
  }\href {\doibase 10.1088/1361-6382/ab5c9a} {\bibfield  {journal} {\bibinfo
  {journal} {Class. Quant. Grav.}\ }\textbf {\bibinfo {volume} {37}},\ \bibinfo
  {pages} {02LT01} (\bibinfo {year} {2020})},\ \Eprint
  {http://arxiv.org/abs/1905.13155} {arXiv:1905.13155 [gr-qc]} \BibitemShut
  {NoStop}%
\bibitem [{\citenamefont {Vincent}\ \emph {et~al.}(2020)\citenamefont
  {Vincent}, \citenamefont {Wielgus}, \citenamefont {Abramowicz}, \citenamefont
  {Gourgoulhon}, \citenamefont {Lasota}, \citenamefont {Paumard},\ and\
  \citenamefont {Perrin}}]{Vincent:2020dij}%
  \BibitemOpen
  \bibfield  {author} {\bibinfo {author} {\bibfnamefont {F.}~\bibnamefont
  {Vincent}}, \bibinfo {author} {\bibfnamefont {M.}~\bibnamefont {Wielgus}},
  \bibinfo {author} {\bibfnamefont {M.}~\bibnamefont {Abramowicz}}, \bibinfo
  {author} {\bibfnamefont {E.}~\bibnamefont {Gourgoulhon}}, \bibinfo {author}
  {\bibfnamefont {J.-P.}\ \bibnamefont {Lasota}}, \bibinfo {author}
  {\bibfnamefont {T.}~\bibnamefont {Paumard}}, \ and\ \bibinfo {author}
  {\bibfnamefont {G.}~\bibnamefont {Perrin}},\ }\href@noop {} {\  (\bibinfo
  {year} {2020})},\ \Eprint {http://arxiv.org/abs/2002.09226} {arXiv:2002.09226
  [gr-qc]} \BibitemShut {NoStop}%
\bibitem [{\citenamefont {Akiyama}\ \emph
  {et~al.}(2019{\natexlab{a}})\citenamefont {Akiyama} \emph
  {et~al.}}]{Akiyama:2019eap}%
  \BibitemOpen
  \bibfield  {author} {\bibinfo {author} {\bibfnamefont {K.}~\bibnamefont
  {Akiyama}} \emph {et~al.} (\bibinfo {collaboration} {Event Horizon
  Telescope}),\ }\href {\doibase 10.3847/2041-8213/ab1141} {\bibfield
  {journal} {\bibinfo  {journal} {Astrophys. J. Lett.}\ }\textbf {\bibinfo
  {volume} {875}},\ \bibinfo {pages} {L6} (\bibinfo {year}
  {2019}{\natexlab{a}})},\ \Eprint {http://arxiv.org/abs/1906.11243}
  {arXiv:1906.11243 [astro-ph.GA]} \BibitemShut {NoStop}%
\bibitem [{\citenamefont {Abdikamalov}\ \emph
  {et~al.}(2019{\natexlab{a}})\citenamefont {Abdikamalov}, \citenamefont
  {Ayzenberg}, \citenamefont {Bambi}, \citenamefont {Nampalliwar},
  \citenamefont {Tripathi}, \citenamefont {Wong}, \citenamefont {Xu},
  \citenamefont {Yan}, \citenamefont {Yan},\ and\ \citenamefont
  {Yang}}]{Abdikamalov:2019zfz}%
  \BibitemOpen
  \bibfield  {author} {\bibinfo {author} {\bibfnamefont {A.~B.}\ \bibnamefont
  {Abdikamalov}}, \bibinfo {author} {\bibfnamefont {D.}~\bibnamefont
  {Ayzenberg}}, \bibinfo {author} {\bibfnamefont {C.}~\bibnamefont {Bambi}},
  \bibinfo {author} {\bibfnamefont {S.}~\bibnamefont {Nampalliwar}}, \bibinfo
  {author} {\bibfnamefont {A.}~\bibnamefont {Tripathi}}, \bibinfo {author}
  {\bibfnamefont {J.}~\bibnamefont {Wong}}, \bibinfo {author} {\bibfnamefont
  {Y.}~\bibnamefont {Xu}}, \bibinfo {author} {\bibfnamefont {J.}~\bibnamefont
  {Yan}}, \bibinfo {author} {\bibfnamefont {Y.}~\bibnamefont {Yan}}, \ and\
  \bibinfo {author} {\bibfnamefont {Y.}~\bibnamefont {Yang}},\ }\href {\doibase
  10.3390/proceedings2019017002} {\bibfield  {journal} {\bibinfo  {journal}
  {MDPI Proc.}\ }\textbf {\bibinfo {volume} {17}},\ \bibinfo {pages} {2}
  (\bibinfo {year} {2019}{\natexlab{a}})},\ \Eprint
  {http://arxiv.org/abs/1905.08012} {arXiv:1905.08012 [gr-qc]} \BibitemShut
  {NoStop}%
\bibitem [{\citenamefont {Penrose}(1965)}]{Penrose:1964wq}%
  \BibitemOpen
  \bibfield  {author} {\bibinfo {author} {\bibfnamefont {R.}~\bibnamefont
  {Penrose}},\ }\href {\doibase 10.1103/PhysRevLett.14.57} {\bibfield
  {journal} {\bibinfo  {journal} {Phys. Rev. Lett.}\ }\textbf {\bibinfo
  {volume} {14}},\ \bibinfo {pages} {57} (\bibinfo {year} {1965})}\BibitemShut
  {NoStop}%
\bibitem [{\citenamefont {'t~Hooft}\ and\ \citenamefont
  {Veltman}(1974)}]{tHooft:1974toh}%
  \BibitemOpen
  \bibfield  {author} {\bibinfo {author} {\bibfnamefont {G.}~\bibnamefont
  {'t~Hooft}}\ and\ \bibinfo {author} {\bibfnamefont {M.}~\bibnamefont
  {Veltman}},\ }\href@noop {} {\bibfield  {journal} {\bibinfo  {journal} {Ann.
  Inst. H. Poincare Phys. Theor. A}\ }\textbf {\bibinfo {volume} {20}},\
  \bibinfo {pages} {69} (\bibinfo {year} {1974})}\BibitemShut {NoStop}%
\bibitem [{\citenamefont {Aharony}\ \emph {et~al.}(2000)\citenamefont
  {Aharony}, \citenamefont {Gubser}, \citenamefont {Maldacena}, \citenamefont
  {Ooguri},\ and\ \citenamefont {Oz}}]{Aharony:1999ti}%
  \BibitemOpen
  \bibfield  {author} {\bibinfo {author} {\bibfnamefont {O.}~\bibnamefont
  {Aharony}}, \bibinfo {author} {\bibfnamefont {S.~S.}\ \bibnamefont {Gubser}},
  \bibinfo {author} {\bibfnamefont {J.~M.}\ \bibnamefont {Maldacena}}, \bibinfo
  {author} {\bibfnamefont {H.}~\bibnamefont {Ooguri}}, \ and\ \bibinfo {author}
  {\bibfnamefont {Y.}~\bibnamefont {Oz}},\ }\href {\doibase
  10.1016/S0370-1573(99)00083-6} {\bibfield  {journal} {\bibinfo  {journal}
  {Phys. Rept.}\ }\textbf {\bibinfo {volume} {323}},\ \bibinfo {pages} {183}
  (\bibinfo {year} {2000})},\ \Eprint {http://arxiv.org/abs/hep-th/9905111}
  {arXiv:hep-th/9905111} \BibitemShut {NoStop}%
\bibitem [{\citenamefont {Ashtekar}\ and\ \citenamefont
  {Singh}(2011)}]{Ashtekar:2011ni}%
  \BibitemOpen
  \bibfield  {author} {\bibinfo {author} {\bibfnamefont {A.}~\bibnamefont
  {Ashtekar}}\ and\ \bibinfo {author} {\bibfnamefont {P.}~\bibnamefont
  {Singh}},\ }\href {\doibase 10.1088/0264-9381/28/21/213001} {\bibfield
  {journal} {\bibinfo  {journal} {Class. Quant. Grav.}\ }\textbf {\bibinfo
  {volume} {28}},\ \bibinfo {pages} {213001} (\bibinfo {year} {2011})},\
  \Eprint {http://arxiv.org/abs/1108.0893} {arXiv:1108.0893 [gr-qc]}
  \BibitemShut {NoStop}%
\bibitem [{\citenamefont {van~de Ven}(1992)}]{vandeVen:1991gw}%
  \BibitemOpen
  \bibfield  {author} {\bibinfo {author} {\bibfnamefont {A.}~\bibnamefont
  {van~de Ven}},\ }\href {\doibase 10.1016/0550-3213(92)90011-Y} {\bibfield
  {journal} {\bibinfo  {journal} {Nucl. Phys. B}\ }\textbf {\bibinfo {volume}
  {378}},\ \bibinfo {pages} {309} (\bibinfo {year} {1992})}\BibitemShut
  {NoStop}%
\bibitem [{\citenamefont {Bonanno}\ and\ \citenamefont
  {Reuter}(2000)}]{Bonanno:2000ep}%
  \BibitemOpen
  \bibfield  {author} {\bibinfo {author} {\bibfnamefont {A.}~\bibnamefont
  {Bonanno}}\ and\ \bibinfo {author} {\bibfnamefont {M.}~\bibnamefont
  {Reuter}},\ }\href {\doibase 10.1103/PhysRevD.62.043008} {\bibfield
  {journal} {\bibinfo  {journal} {Phys. Rev. D}\ }\textbf {\bibinfo {volume}
  {62}},\ \bibinfo {pages} {043008} (\bibinfo {year} {2000})},\ \Eprint
  {http://arxiv.org/abs/hep-th/0002196} {arXiv:hep-th/0002196} \BibitemShut
  {NoStop}%
\bibitem [{\citenamefont {Reuter}\ and\ \citenamefont
  {Saueressig}(2012)}]{Reuter:2012id}%
  \BibitemOpen
  \bibfield  {author} {\bibinfo {author} {\bibfnamefont {M.}~\bibnamefont
  {Reuter}}\ and\ \bibinfo {author} {\bibfnamefont {F.}~\bibnamefont
  {Saueressig}},\ }\href {\doibase 10.1088/1367-2630/14/5/055022} {\bibfield
  {journal} {\bibinfo  {journal} {New J. Phys.}\ }\textbf {\bibinfo {volume}
  {14}},\ \bibinfo {pages} {055022} (\bibinfo {year} {2012})},\ \Eprint
  {http://arxiv.org/abs/1202.2274} {arXiv:1202.2274 [hep-th]} \BibitemShut
  {NoStop}%
\bibitem [{\citenamefont {Bonanno}\ \emph {et~al.}(2020)\citenamefont
  {Bonanno}, \citenamefont {Eichhorn}, \citenamefont {Gies}, \citenamefont
  {Pawlowski}, \citenamefont {Percacci}, \citenamefont {Reuter}, \citenamefont
  {Saueressig},\ and\ \citenamefont {Vacca}}]{Bonanno:2020bil}%
  \BibitemOpen
  \bibfield  {author} {\bibinfo {author} {\bibfnamefont {A.}~\bibnamefont
  {Bonanno}}, \bibinfo {author} {\bibfnamefont {A.}~\bibnamefont {Eichhorn}},
  \bibinfo {author} {\bibfnamefont {H.}~\bibnamefont {Gies}}, \bibinfo {author}
  {\bibfnamefont {J.~M.}\ \bibnamefont {Pawlowski}}, \bibinfo {author}
  {\bibfnamefont {R.}~\bibnamefont {Percacci}}, \bibinfo {author}
  {\bibfnamefont {M.}~\bibnamefont {Reuter}}, \bibinfo {author} {\bibfnamefont
  {F.}~\bibnamefont {Saueressig}}, \ and\ \bibinfo {author} {\bibfnamefont
  {G.~P.}\ \bibnamefont {Vacca}},\ }\href {\doibase 10.3389/fphy.2020.00269}
  {\bibfield  {journal} {\bibinfo  {journal} {Front. in Phys.}\ }\textbf
  {\bibinfo {volume} {8}},\ \bibinfo {pages} {269} (\bibinfo {year} {2020})},\
  \Eprint {http://arxiv.org/abs/2004.06810} {arXiv:2004.06810 [gr-qc]}
  \BibitemShut {NoStop}%
\bibitem [{\citenamefont {Weinberg}(1996)}]{Weinberg:1996kw}%
  \BibitemOpen
  \bibfield  {author} {\bibinfo {author} {\bibfnamefont {S.}~\bibnamefont
  {Weinberg}},\ }in\ \href@noop {} {\emph {\bibinfo {booktitle} {{Conference on
  Historical Examination and Philosophical Reflections on the Foundations of
  Quantum Field Theory}}}}\ (\bibinfo {year} {1996})\ pp.\ \bibinfo {pages}
  {241--251},\ \Eprint {http://arxiv.org/abs/hep-th/9702027}
  {arXiv:hep-th/9702027} \BibitemShut {NoStop}%
\bibitem [{\citenamefont {Weinberg}(1980)}]{Weinberg:1980gg}%
  \BibitemOpen
  \bibfield  {author} {\bibinfo {author} {\bibfnamefont {S.}~\bibnamefont
  {Weinberg}},\ }\enquote {\bibinfo {title} {{ULTRAVIOLET DIVERGENCES IN
  QUANTUM THEORIES OF GRAVITATION}},}\ in\ \href@noop {} {\emph {\bibinfo
  {booktitle} {{General Relativity}: {An Einstein Centenary Survey}}}}\
  (\bibinfo {year} {1980})\ pp.\ \bibinfo {pages} {790--831}\BibitemShut
  {NoStop}%
\bibitem [{\citenamefont {Held}\ \emph {et~al.}(2019)\citenamefont {Held},
  \citenamefont {Gold},\ and\ \citenamefont {Eichhorn}}]{Held:2019xde}%
  \BibitemOpen
  \bibfield  {author} {\bibinfo {author} {\bibfnamefont {A.}~\bibnamefont
  {Held}}, \bibinfo {author} {\bibfnamefont {R.}~\bibnamefont {Gold}}, \ and\
  \bibinfo {author} {\bibfnamefont {A.}~\bibnamefont {Eichhorn}},\ }\href
  {\doibase 10.1088/1475-7516/2019/06/029} {\bibfield  {journal} {\bibinfo
  {journal} {JCAP}\ }\textbf {\bibinfo {volume} {06}},\ \bibinfo {pages} {029}
  (\bibinfo {year} {2019})},\ \Eprint {http://arxiv.org/abs/1904.07133}
  {arXiv:1904.07133 [gr-qc]} \BibitemShut {NoStop}%
\bibitem [{\citenamefont {Elbert}\ \emph {et~al.}(2018)\citenamefont {Elbert},
  \citenamefont {Bullock},\ and\ \citenamefont {Kaplinghat}}]{Elbert:2017sbr}%
  \BibitemOpen
  \bibfield  {author} {\bibinfo {author} {\bibfnamefont {O.~D.}\ \bibnamefont
  {Elbert}}, \bibinfo {author} {\bibfnamefont {J.~S.}\ \bibnamefont {Bullock}},
  \ and\ \bibinfo {author} {\bibfnamefont {M.}~\bibnamefont {Kaplinghat}},\
  }\href {\doibase 10.1093/mnras/stx1959} {\bibfield  {journal} {\bibinfo
  {journal} {Mon. Not. Roy. Astron. Soc.}\ }\textbf {\bibinfo {volume} {473}},\
  \bibinfo {pages} {1186} (\bibinfo {year} {2018})},\ \Eprint
  {http://arxiv.org/abs/1703.02551} {arXiv:1703.02551 [astro-ph.GA]}
  \BibitemShut {NoStop}%
\bibitem [{\citenamefont {Berti}(2019)}]{Berti:2019tcy}%
  \BibitemOpen
  \bibfield  {author} {\bibinfo {author} {\bibfnamefont {E.}~\bibnamefont
  {Berti}},\ }\href {\doibase 10.1007/s10714-019-2622-2} {\bibfield  {journal}
  {\bibinfo  {journal} {Gen. Rel. Grav.}\ }\textbf {\bibinfo {volume} {51}},\
  \bibinfo {pages} {140} (\bibinfo {year} {2019})},\ \Eprint
  {http://arxiv.org/abs/1911.00541} {arXiv:1911.00541 [gr-qc]} \BibitemShut
  {NoStop}%
\bibitem [{\citenamefont {Cardenas-Avendano}\ \emph {et~al.}(2019)\citenamefont
  {Cardenas-Avendano}, \citenamefont {Nampalliwar},\ and\ \citenamefont
  {Yunes}}]{CardenasAvendano:2019zxd}%
  \BibitemOpen
  \bibfield  {author} {\bibinfo {author} {\bibfnamefont {A.}~\bibnamefont
  {Cardenas-Avendano}}, \bibinfo {author} {\bibfnamefont {S.}~\bibnamefont
  {Nampalliwar}}, \ and\ \bibinfo {author} {\bibfnamefont {N.}~\bibnamefont
  {Yunes}},\ }\href@noop {} {\  (\bibinfo {year} {2019})},\ \Eprint
  {http://arxiv.org/abs/1912.08062} {arXiv:1912.08062 [gr-qc]} \BibitemShut
  {NoStop}%
\bibitem [{\citenamefont {Barausse}\ \emph {et~al.}(2020)\citenamefont
  {Barausse} \emph {et~al.}}]{Barausse:2020rsu}%
  \BibitemOpen
  \bibfield  {author} {\bibinfo {author} {\bibfnamefont {E.}~\bibnamefont
  {Barausse}} \emph {et~al.},\ }\href {\doibase 10.1007/s10714-020-02691-1}
  {\bibfield  {journal} {\bibinfo  {journal} {Gen. Rel. Grav.}\ }\textbf
  {\bibinfo {volume} {52}},\ \bibinfo {pages} {81} (\bibinfo {year} {2020})},\
  \Eprint {http://arxiv.org/abs/2001.09793} {arXiv:2001.09793 [gr-qc]}
  \BibitemShut {NoStop}%
\bibitem [{\citenamefont {Bambi}(2014)}]{Bambi:2013mha}%
  \BibitemOpen
  \bibfield  {author} {\bibinfo {author} {\bibfnamefont {C.}~\bibnamefont
  {Bambi}},\ }\href {\doibase 10.1088/1475-7516/2014/03/034} {\bibfield
  {journal} {\bibinfo  {journal} {JCAP}\ }\textbf {\bibinfo {volume} {03}},\
  \bibinfo {pages} {034} (\bibinfo {year} {2014})},\ \Eprint
  {http://arxiv.org/abs/1308.2470} {arXiv:1308.2470 [gr-qc]} \BibitemShut
  {NoStop}%
\bibitem [{\citenamefont {Li}\ \emph {et~al.}(2019)\citenamefont {Li},
  \citenamefont {Yan}, \citenamefont {Xue}, \citenamefont {Ren}, \citenamefont
  {Cai}, \citenamefont {Easson}, \citenamefont {Yuan},\ and\ \citenamefont
  {Zhao}}]{Li:2019lsm}%
  \BibitemOpen
  \bibfield  {author} {\bibinfo {author} {\bibfnamefont {C.}~\bibnamefont
  {Li}}, \bibinfo {author} {\bibfnamefont {S.-F.}\ \bibnamefont {Yan}},
  \bibinfo {author} {\bibfnamefont {L.}~\bibnamefont {Xue}}, \bibinfo {author}
  {\bibfnamefont {X.}~\bibnamefont {Ren}}, \bibinfo {author} {\bibfnamefont
  {Y.-F.}\ \bibnamefont {Cai}}, \bibinfo {author} {\bibfnamefont {D.~A.}\
  \bibnamefont {Easson}}, \bibinfo {author} {\bibfnamefont {Y.-F.}\
  \bibnamefont {Yuan}}, \ and\ \bibinfo {author} {\bibfnamefont
  {H.}~\bibnamefont {Zhao}},\ }\href@noop {} {\  (\bibinfo {year} {2019})},\
  \Eprint {http://arxiv.org/abs/1912.12629} {arXiv:1912.12629 [astro-ph.CO]}
  \BibitemShut {NoStop}%
\bibitem [{\citenamefont {Reynolds}(2014)}]{Reynolds:2013qqa}%
  \BibitemOpen
  \bibfield  {author} {\bibinfo {author} {\bibfnamefont {C.~S.}\ \bibnamefont
  {Reynolds}},\ }\href {\doibase 10.1007/s11214-013-0006-6} {\bibfield
  {journal} {\bibinfo  {journal} {Space Sci. Rev.}\ }\textbf {\bibinfo {volume}
  {183}},\ \bibinfo {pages} {277} (\bibinfo {year} {2014})},\ \Eprint
  {http://arxiv.org/abs/1302.3260} {arXiv:1302.3260 [astro-ph.HE]} \BibitemShut
  {NoStop}%
\bibitem [{\citenamefont {Bambi}\ \emph {et~al.}(2018)\citenamefont {Bambi},
  \citenamefont {Abdikamalov}, \citenamefont {Ayzenberg}, \citenamefont {Cao},
  \citenamefont {Liu}, \citenamefont {Nampalliwar}, \citenamefont {Tripathi},
  \citenamefont {Wang-Ji},\ and\ \citenamefont {Xu}}]{Bambi2018}%
  \BibitemOpen
  \bibfield  {author} {\bibinfo {author} {\bibfnamefont {C.}~\bibnamefont
  {Bambi}}, \bibinfo {author} {\bibfnamefont {A.~B.}\ \bibnamefont
  {Abdikamalov}}, \bibinfo {author} {\bibfnamefont {D.}~\bibnamefont
  {Ayzenberg}}, \bibinfo {author} {\bibfnamefont {Z.}~\bibnamefont {Cao}},
  \bibinfo {author} {\bibfnamefont {H.}~\bibnamefont {Liu}}, \bibinfo {author}
  {\bibfnamefont {S.}~\bibnamefont {Nampalliwar}}, \bibinfo {author}
  {\bibfnamefont {A.}~\bibnamefont {Tripathi}}, \bibinfo {author}
  {\bibfnamefont {J.}~\bibnamefont {Wang-Ji}}, \ and\ \bibinfo {author}
  {\bibfnamefont {Y.}~\bibnamefont {Xu}},\ }\href {\doibase
  10.3390/universe4070079} {\bibfield  {journal} {\bibinfo  {journal}
  {Universe}\ }\textbf {\bibinfo {volume} {4}},\ \bibinfo {pages} {79}
  (\bibinfo {year} {2018})},\ \Eprint {http://arxiv.org/abs/1806.02141}
  {arXiv:1806.02141 [gr-qc]} \BibitemShut {NoStop}%
\bibitem [{\citenamefont {Mizuno}\ \emph {et~al.}(2018)\citenamefont {Mizuno},
  \citenamefont {Younsi}, \citenamefont {Fromm}, \citenamefont {Porth},
  \citenamefont {De~Laurentis}, \citenamefont {Olivares}, \citenamefont
  {Falcke}, \citenamefont {Kramer},\ and\ \citenamefont
  {Rezzolla}}]{Mizuno:2018lxz}%
  \BibitemOpen
  \bibfield  {author} {\bibinfo {author} {\bibfnamefont {Y.}~\bibnamefont
  {Mizuno}}, \bibinfo {author} {\bibfnamefont {Z.}~\bibnamefont {Younsi}},
  \bibinfo {author} {\bibfnamefont {C.~M.}\ \bibnamefont {Fromm}}, \bibinfo
  {author} {\bibfnamefont {O.}~\bibnamefont {Porth}}, \bibinfo {author}
  {\bibfnamefont {M.}~\bibnamefont {De~Laurentis}}, \bibinfo {author}
  {\bibfnamefont {H.}~\bibnamefont {Olivares}}, \bibinfo {author}
  {\bibfnamefont {H.}~\bibnamefont {Falcke}}, \bibinfo {author} {\bibfnamefont
  {M.}~\bibnamefont {Kramer}}, \ and\ \bibinfo {author} {\bibfnamefont
  {L.}~\bibnamefont {Rezzolla}},\ }\href {\doibase 10.1038/s41550-018-0449-5}
  {\bibfield  {journal} {\bibinfo  {journal} {Nat. Astron.}\ }\textbf {\bibinfo
  {volume} {2}},\ \bibinfo {pages} {585} (\bibinfo {year} {2018})},\ \Eprint
  {http://arxiv.org/abs/1804.05812} {arXiv:1804.05812 [astro-ph.GA]}
  \BibitemShut {NoStop}%
\bibitem [{\citenamefont {Chrusciel}\ \emph {et~al.}(2012)\citenamefont
  {Chrusciel}, \citenamefont {Lopes~Costa},\ and\ \citenamefont
  {Heusler}}]{Chrusciel2012}%
  \BibitemOpen
  \bibfield  {author} {\bibinfo {author} {\bibfnamefont {P.~T.}\ \bibnamefont
  {Chrusciel}}, \bibinfo {author} {\bibfnamefont {J.}~\bibnamefont
  {Lopes~Costa}}, \ and\ \bibinfo {author} {\bibfnamefont {M.}~\bibnamefont
  {Heusler}},\ }\href {\doibase 10.12942/lrr-2012-7} {\bibfield  {journal}
  {\bibinfo  {journal} {Living Rev. Rel.}\ }\textbf {\bibinfo {volume} {15}},\
  \bibinfo {pages} {7} (\bibinfo {year} {2012})},\ \Eprint
  {http://arxiv.org/abs/1205.6112} {arXiv:1205.6112 [gr-qc]} \BibitemShut
  {NoStop}%
\bibitem [{\citenamefont {Bambi}\ \emph {et~al.}(2009)\citenamefont {Bambi},
  \citenamefont {Dolgov},\ and\ \citenamefont {Petrov}}]{Bambi:2008hp}%
  \BibitemOpen
  \bibfield  {author} {\bibinfo {author} {\bibfnamefont {C.}~\bibnamefont
  {Bambi}}, \bibinfo {author} {\bibfnamefont {A.~D.}\ \bibnamefont {Dolgov}}, \
  and\ \bibinfo {author} {\bibfnamefont {A.~A.}\ \bibnamefont {Petrov}},\
  }\href {\doibase 10.1088/1475-7516/2009/09/013} {\bibfield  {journal}
  {\bibinfo  {journal} {JCAP}\ }\textbf {\bibinfo {volume} {09}},\ \bibinfo
  {pages} {013} (\bibinfo {year} {2009})},\ \Eprint
  {http://arxiv.org/abs/0806.3440} {arXiv:0806.3440 [astro-ph]} \BibitemShut
  {NoStop}%
\bibitem [{\citenamefont {Carter}(1971)}]{Carter1971}%
  \BibitemOpen
  \bibfield  {author} {\bibinfo {author} {\bibfnamefont {B.}~\bibnamefont
  {Carter}},\ }\href {\doibase 10.1103/PhysRevLett.26.331} {\bibfield
  {journal} {\bibinfo  {journal} {Phys. Rev. Lett.}\ }\textbf {\bibinfo
  {volume} {26}},\ \bibinfo {pages} {331} (\bibinfo {year} {1971})}\BibitemShut
  {NoStop}%
\bibitem [{\citenamefont {Robinson}(1975)}]{Robinson1975}%
  \BibitemOpen
  \bibfield  {author} {\bibinfo {author} {\bibfnamefont {D.~C.}\ \bibnamefont
  {Robinson}},\ }\href {\doibase 10.1103/PhysRevLett.34.905} {\bibfield
  {journal} {\bibinfo  {journal} {Phys. Rev. Lett.}\ }\textbf {\bibinfo
  {volume} {34}},\ \bibinfo {pages} {905} (\bibinfo {year} {1975})}\BibitemShut
  {NoStop}%
\bibitem [{\citenamefont {Reuter}\ and\ \citenamefont
  {Tuiran}(2011)}]{Reuter:2010xb}%
  \BibitemOpen
  \bibfield  {author} {\bibinfo {author} {\bibfnamefont {M.}~\bibnamefont
  {Reuter}}\ and\ \bibinfo {author} {\bibfnamefont {E.}~\bibnamefont
  {Tuiran}},\ }\href {\doibase 10.1103/PhysRevD.83.044041} {\bibfield
  {journal} {\bibinfo  {journal} {Phys. Rev. D}\ }\textbf {\bibinfo {volume}
  {83}},\ \bibinfo {pages} {044041} (\bibinfo {year} {2011})},\ \Eprint
  {http://arxiv.org/abs/1009.3528} {arXiv:1009.3528 [hep-th]} \BibitemShut
  {NoStop}%
\bibitem [{\citenamefont {Koch}\ and\ \citenamefont
  {Saueressig}(2014)}]{Koch:2014cqa}%
  \BibitemOpen
  \bibfield  {author} {\bibinfo {author} {\bibfnamefont {B.}~\bibnamefont
  {Koch}}\ and\ \bibinfo {author} {\bibfnamefont {F.}~\bibnamefont
  {Saueressig}},\ }\href {\doibase 10.1142/S0217751X14300117} {\bibfield
  {journal} {\bibinfo  {journal} {Int. J. Mod. Phys. A}\ }\textbf {\bibinfo
  {volume} {29}},\ \bibinfo {pages} {1430011} (\bibinfo {year} {2014})},\
  \Eprint {http://arxiv.org/abs/1401.4452} {arXiv:1401.4452 [hep-th]}
  \BibitemShut {NoStop}%
\bibitem [{\citenamefont {Platania}(2019)}]{Platania:2019kyx}%
  \BibitemOpen
  \bibfield  {author} {\bibinfo {author} {\bibfnamefont {A.}~\bibnamefont
  {Platania}},\ }\href {\doibase 10.1140/epjc/s10052-019-6990-2} {\bibfield
  {journal} {\bibinfo  {journal} {Eur. Phys. J. C}\ }\textbf {\bibinfo {volume}
  {79}},\ \bibinfo {pages} {470} (\bibinfo {year} {2019})},\ \Eprint
  {http://arxiv.org/abs/1903.10411} {arXiv:1903.10411 [gr-qc]} \BibitemShut
  {NoStop}%
\bibitem [{\citenamefont {Rinc\'on}\ and\ \citenamefont
  {Panotopoulos}(2020)}]{Rincon:2020iwy}%
  \BibitemOpen
  \bibfield  {author} {\bibinfo {author} {\bibfnamefont {A.}~\bibnamefont
  {Rinc\'on}}\ and\ \bibinfo {author} {\bibfnamefont {G.}~\bibnamefont
  {Panotopoulos}},\ }\href {\doibase 10.1016/j.dark.2020.100639} {\bibfield
  {journal} {\bibinfo  {journal} {Phys. Dark Univ.}\ }\textbf {\bibinfo
  {volume} {30}},\ \bibinfo {pages} {100639} (\bibinfo {year} {2020})},\
  \Eprint {http://arxiv.org/abs/2006.11889} {arXiv:2006.11889 [gr-qc]}
  \BibitemShut {NoStop}%
\bibitem [{\citenamefont {Akiyama}\ \emph
  {et~al.}(2019{\natexlab{b}})\citenamefont {Akiyama} \emph
  {et~al.}}]{Akiyama:2019cqa}%
  \BibitemOpen
  \bibfield  {author} {\bibinfo {author} {\bibfnamefont {K.}~\bibnamefont
  {Akiyama}} \emph {et~al.} (\bibinfo {collaboration} {Event Horizon
  Telescope}),\ }\href {\doibase 10.3847/2041-8213/ab0ec7} {\bibfield
  {journal} {\bibinfo  {journal} {Astrophys. J.}\ }\textbf {\bibinfo {volume}
  {875}},\ \bibinfo {pages} {L1} (\bibinfo {year} {2019}{\natexlab{b}})},\
  \Eprint {http://arxiv.org/abs/1906.11238} {arXiv:1906.11238 [astro-ph.GA]}
  \BibitemShut {NoStop}%
\bibitem [{\citenamefont {Fabian}\ \emph {et~al.}(1989)\citenamefont {Fabian},
  \citenamefont {Rees}, \citenamefont {Stella},\ and\ \citenamefont
  {White}}]{Fabian:1989ej}%
  \BibitemOpen
  \bibfield  {author} {\bibinfo {author} {\bibfnamefont {A.}~\bibnamefont
  {Fabian}}, \bibinfo {author} {\bibfnamefont {M.}~\bibnamefont {Rees}},
  \bibinfo {author} {\bibfnamefont {L.}~\bibnamefont {Stella}}, \ and\ \bibinfo
  {author} {\bibfnamefont {N.}~\bibnamefont {White}},\ }\href@noop {}
  {\bibfield  {journal} {\bibinfo  {journal} {Mon. Not. Roy. Astron. Soc.}\
  }\textbf {\bibinfo {volume} {238}},\ \bibinfo {pages} {729} (\bibinfo {year}
  {1989})}\BibitemShut {NoStop}%
\bibitem [{\citenamefont {Laor}(1991)}]{Laor:1991nc}%
  \BibitemOpen
  \bibfield  {author} {\bibinfo {author} {\bibfnamefont {A.}~\bibnamefont
  {Laor}},\ }\href {\doibase 10.1086/170257} {\bibfield  {journal} {\bibinfo
  {journal} {Astrophys. J.}\ }\textbf {\bibinfo {volume} {376}},\ \bibinfo
  {pages} {90} (\bibinfo {year} {1991})}\BibitemShut {NoStop}%
\bibitem [{\citenamefont {Reynolds}(2019)}]{Reynolds:2019uxi}%
  \BibitemOpen
  \bibfield  {author} {\bibinfo {author} {\bibfnamefont {C.~S.}\ \bibnamefont
  {Reynolds}},\ }\href {\doibase 10.1038/s41550-018-0665-z} {\bibfield
  {journal} {\bibinfo  {journal} {Nature Astron.}\ }\textbf {\bibinfo {volume}
  {3}},\ \bibinfo {pages} {41} (\bibinfo {year} {2019})},\ \Eprint
  {http://arxiv.org/abs/1903.11704} {arXiv:1903.11704 [astro-ph.HE]}
  \BibitemShut {NoStop}%
\bibitem [{\citenamefont {Garc\'{i}a}\ \emph {et~al.}(2014)\citenamefont
  {Garc\'{i}a} \emph {et~al.}}]{Garcia2013}%
  \BibitemOpen
  \bibfield  {author} {\bibinfo {author} {\bibfnamefont {J.}~\bibnamefont
  {Garc\'{i}a}} \emph {et~al.},\ }\href {\doibase 10.1088/0004-637X/782/2/76}
  {\bibfield  {journal} {\bibinfo  {journal} {Astrophys. J.}\ }\textbf
  {\bibinfo {volume} {782}},\ \bibinfo {pages} {76} (\bibinfo {year} {2014})},\
  \Eprint {http://arxiv.org/abs/1312.3231} {arXiv:1312.3231 [astro-ph.HE]}
  \BibitemShut {NoStop}%
\bibitem [{\citenamefont {Dauser}\ \emph {et~al.}(2014)\citenamefont {Dauser}
  \emph {et~al.}}]{Dauser2014}%
  \BibitemOpen
  \bibfield  {author} {\bibinfo {author} {\bibfnamefont {T.}~\bibnamefont
  {Dauser}} \emph {et~al.},\ }\href {\doibase 10.1093/mnrasl/slu125} {\bibfield
   {journal} {\bibinfo  {journal} {Mon. Not. Roy. Astron. Soc.}\ }\textbf
  {\bibinfo {volume} {444}},\ \bibinfo {pages} {100} (\bibinfo {year}
  {2014})},\ \Eprint {http://arxiv.org/abs/1408.2347} {arXiv:1408.2347
  [astro-ph.HE]} \BibitemShut {NoStop}%
\bibitem [{\citenamefont {Bambi}\ \emph
  {et~al.}(2017{\natexlab{a}})\citenamefont {Bambi} \emph
  {et~al.}}]{relxillnk}%
  \BibitemOpen
  \bibfield  {author} {\bibinfo {author} {\bibfnamefont {C.}~\bibnamefont
  {Bambi}} \emph {et~al.},\ }\href {\doibase 10.3847/1538-4357/aa74c0}
  {\bibfield  {journal} {\bibinfo  {journal} {Astrophys. J.}\ }\textbf
  {\bibinfo {volume} {842}},\ \bibinfo {pages} {76} (\bibinfo {year}
  {2017}{\natexlab{a}})},\ \Eprint {http://arxiv.org/abs/1607.00596}
  {arXiv:1607.00596 [gr-qc]} \BibitemShut {NoStop}%
\bibitem [{\citenamefont {Abdikamalov}\ \emph
  {et~al.}(2019{\natexlab{b}})\citenamefont {Abdikamalov}, \citenamefont
  {Ayzenberg}, \citenamefont {Bambi}, \citenamefont {Dauser}, \citenamefont
  {Garcia},\ and\ \citenamefont {Nampalliwar}}]{Abdikamalov:2019yrr}%
  \BibitemOpen
  \bibfield  {author} {\bibinfo {author} {\bibfnamefont {A.~B.}\ \bibnamefont
  {Abdikamalov}}, \bibinfo {author} {\bibfnamefont {D.}~\bibnamefont
  {Ayzenberg}}, \bibinfo {author} {\bibfnamefont {C.}~\bibnamefont {Bambi}},
  \bibinfo {author} {\bibfnamefont {T.}~\bibnamefont {Dauser}}, \bibinfo
  {author} {\bibfnamefont {J.~A.}\ \bibnamefont {Garcia}}, \ and\ \bibinfo
  {author} {\bibfnamefont {S.}~\bibnamefont {Nampalliwar}},\ }\href {\doibase
  10.3847/1538-4357/ab1f89} {\bibfield  {journal} {\bibinfo  {journal}
  {Astrophys. J.}\ }\textbf {\bibinfo {volume} {878}},\ \bibinfo {pages} {91}
  (\bibinfo {year} {2019}{\natexlab{b}})},\ \Eprint
  {http://arxiv.org/abs/1902.09665} {arXiv:1902.09665 [gr-qc]} \BibitemShut
  {NoStop}%
\bibitem [{\citenamefont {Bambi}\ \emph
  {et~al.}(2017{\natexlab{b}})\citenamefont {Bambi}, \citenamefont {Modesto},\
  and\ \citenamefont {Rachwa\l}}]{Bambi:2016wdn}%
  \BibitemOpen
  \bibfield  {author} {\bibinfo {author} {\bibfnamefont {C.}~\bibnamefont
  {Bambi}}, \bibinfo {author} {\bibfnamefont {L.}~\bibnamefont {Modesto}}, \
  and\ \bibinfo {author} {\bibfnamefont {L.}~\bibnamefont {Rachwa\l}},\ }\href
  {\doibase 10.1088/1475-7516/2017/05/003} {\bibfield  {journal} {\bibinfo
  {journal} {JCAP}\ }\textbf {\bibinfo {volume} {05}},\ \bibinfo {pages} {003}
  (\bibinfo {year} {2017}{\natexlab{b}})},\ \Eprint
  {http://arxiv.org/abs/1611.00865} {arXiv:1611.00865 [gr-qc]} \BibitemShut
  {NoStop}%
\bibitem [{\citenamefont {Horowitz}\ and\ \citenamefont
  {Wiseman}(2012)}]{Horowitz:2011cq}%
  \BibitemOpen
  \bibfield  {author} {\bibinfo {author} {\bibfnamefont {G.~T.}\ \bibnamefont
  {Horowitz}}\ and\ \bibinfo {author} {\bibfnamefont {T.}~\bibnamefont
  {Wiseman}},\ }\enquote {\bibinfo {title} {{General black holes in
  Kaluza--Klein theory}},}\ in\ \href@noop {} {\emph {\bibinfo {booktitle}
  {{Black holes in higher dimensions}}}}\ (\bibinfo {year} {2012})\ pp.\
  \bibinfo {pages} {69--98},\ \Eprint {http://arxiv.org/abs/1107.5563}
  {arXiv:1107.5563 [gr-qc]} \BibitemShut {NoStop}%
\bibitem [{\citenamefont {Johannsen}(2013)}]{Johannsen2015}%
  \BibitemOpen
  \bibfield  {author} {\bibinfo {author} {\bibfnamefont {T.}~\bibnamefont
  {Johannsen}},\ }\href {\doibase 10.1103/PhysRevD.88.044002} {\bibfield
  {journal} {\bibinfo  {journal} {Phys. Rev.}\ }\textbf {\bibinfo {volume}
  {D88}},\ \bibinfo {pages} {044002} (\bibinfo {year} {2013})},\ \Eprint
  {http://arxiv.org/abs/1501.02809} {arXiv:1501.02809 [gr-qc]} \BibitemShut
  {NoStop}%
\bibitem [{\citenamefont {Konoplya}\ \emph {et~al.}(2016)\citenamefont
  {Konoplya}, \citenamefont {Rezzolla},\ and\ \citenamefont
  {Zhidenko}}]{Konoplya2016}%
  \BibitemOpen
  \bibfield  {author} {\bibinfo {author} {\bibfnamefont {R.}~\bibnamefont
  {Konoplya}}, \bibinfo {author} {\bibfnamefont {L.}~\bibnamefont {Rezzolla}},
  \ and\ \bibinfo {author} {\bibfnamefont {A.}~\bibnamefont {Zhidenko}},\
  }\href {\doibase 10.1103/PhysRevD.93.064015} {\bibfield  {journal} {\bibinfo
  {journal} {Phys. Rev.}\ }\textbf {\bibinfo {volume} {D93}},\ \bibinfo {pages}
  {064015} (\bibinfo {year} {2016})},\ \Eprint
  {http://arxiv.org/abs/1602.02378} {arXiv:1602.02378 [gr-qc]} \BibitemShut
  {NoStop}%
\bibitem [{\citenamefont {Nampalliwar}\ \emph {et~al.}(2019)\citenamefont
  {Nampalliwar}, \citenamefont {Xin}, \citenamefont {Srivastava}, \citenamefont
  {Abdikamalov}, \citenamefont {Ayzenberg}, \citenamefont {Bambi},
  \citenamefont {Dauser}, \citenamefont {Garcia},\ and\ \citenamefont
  {Tripathi}}]{Nampalliwar:2019iti}%
  \BibitemOpen
  \bibfield  {author} {\bibinfo {author} {\bibfnamefont {S.}~\bibnamefont
  {Nampalliwar}}, \bibinfo {author} {\bibfnamefont {S.}~\bibnamefont {Xin}},
  \bibinfo {author} {\bibfnamefont {S.}~\bibnamefont {Srivastava}}, \bibinfo
  {author} {\bibfnamefont {A.~B.}\ \bibnamefont {Abdikamalov}}, \bibinfo
  {author} {\bibfnamefont {D.}~\bibnamefont {Ayzenberg}}, \bibinfo {author}
  {\bibfnamefont {C.}~\bibnamefont {Bambi}}, \bibinfo {author} {\bibfnamefont
  {T.}~\bibnamefont {Dauser}}, \bibinfo {author} {\bibfnamefont {J.~A.}\
  \bibnamefont {Garcia}}, \ and\ \bibinfo {author} {\bibfnamefont
  {A.}~\bibnamefont {Tripathi}},\ }\href@noop {} {\  (\bibinfo {year}
  {2019})},\ \Eprint {http://arxiv.org/abs/1903.12119} {arXiv:1903.12119
  [gr-qc]} \BibitemShut {NoStop}%
\bibitem [{\citenamefont {Liu}\ \emph {et~al.}(2019)\citenamefont {Liu},
  \citenamefont {Abdikamalov}, \citenamefont {Ayzenberg}, \citenamefont
  {Bambi}, \citenamefont {Dauser}, \citenamefont {Garcia},\ and\ \citenamefont
  {Nampalliwar}}]{Liu:2019vqh}%
  \BibitemOpen
  \bibfield  {author} {\bibinfo {author} {\bibfnamefont {H.}~\bibnamefont
  {Liu}}, \bibinfo {author} {\bibfnamefont {A.~B.}\ \bibnamefont
  {Abdikamalov}}, \bibinfo {author} {\bibfnamefont {D.}~\bibnamefont
  {Ayzenberg}}, \bibinfo {author} {\bibfnamefont {C.}~\bibnamefont {Bambi}},
  \bibinfo {author} {\bibfnamefont {T.}~\bibnamefont {Dauser}}, \bibinfo
  {author} {\bibfnamefont {J.~A.}\ \bibnamefont {Garcia}}, \ and\ \bibinfo
  {author} {\bibfnamefont {S.}~\bibnamefont {Nampalliwar}},\ }\href {\doibase
  10.1103/PhysRevD.99.123007} {\bibfield  {journal} {\bibinfo  {journal} {Phys.
  Rev. D}\ }\textbf {\bibinfo {volume} {99}},\ \bibinfo {pages} {123007}
  (\bibinfo {year} {2019})},\ \Eprint {http://arxiv.org/abs/1904.08027}
  {arXiv:1904.08027 [gr-qc]} \BibitemShut {NoStop}%
\bibitem [{\citenamefont {Zhou}\ \emph {et~al.}(2020)\citenamefont {Zhou},
  \citenamefont {Tripathi}, \citenamefont {Abdikamalov}, \citenamefont
  {Ayzenberg}, \citenamefont {Bambi}, \citenamefont {Nampalliwar},\ and\
  \citenamefont {Zhou}}]{Zhou:2019kwb}%
  \BibitemOpen
  \bibfield  {author} {\bibinfo {author} {\bibfnamefont {B.}~\bibnamefont
  {Zhou}}, \bibinfo {author} {\bibfnamefont {A.}~\bibnamefont {Tripathi}},
  \bibinfo {author} {\bibfnamefont {A.~B.}\ \bibnamefont {Abdikamalov}},
  \bibinfo {author} {\bibfnamefont {D.}~\bibnamefont {Ayzenberg}}, \bibinfo
  {author} {\bibfnamefont {C.}~\bibnamefont {Bambi}}, \bibinfo {author}
  {\bibfnamefont {S.}~\bibnamefont {Nampalliwar}}, \ and\ \bibinfo {author}
  {\bibfnamefont {M.}~\bibnamefont {Zhou}},\ }\href {\doibase
  10.1140/epjc/s10052-020-7998-3} {\bibfield  {journal} {\bibinfo  {journal}
  {Eur. Phys. J. C}\ }\textbf {\bibinfo {volume} {80}},\ \bibinfo {pages} {400}
  (\bibinfo {year} {2020})},\ \Eprint {http://arxiv.org/abs/1908.05177}
  {arXiv:1908.05177 [gr-qc]} \BibitemShut {NoStop}%
\bibitem [{\citenamefont {Cao}\ \emph {et~al.}(2018)\citenamefont {Cao},
  \citenamefont {Nampalliwar}, \citenamefont {Bambi}, \citenamefont {Dauser},\
  and\ \citenamefont {Garcia}}]{Cao2017}%
  \BibitemOpen
  \bibfield  {author} {\bibinfo {author} {\bibfnamefont {Z.}~\bibnamefont
  {Cao}}, \bibinfo {author} {\bibfnamefont {S.}~\bibnamefont {Nampalliwar}},
  \bibinfo {author} {\bibfnamefont {C.}~\bibnamefont {Bambi}}, \bibinfo
  {author} {\bibfnamefont {T.}~\bibnamefont {Dauser}}, \ and\ \bibinfo {author}
  {\bibfnamefont {J.~A.}\ \bibnamefont {Garcia}},\ }\href {\doibase
  10.1103/PhysRevLett.120.051101} {\bibfield  {journal} {\bibinfo  {journal}
  {Phys. Rev. Lett.}\ }\textbf {\bibinfo {volume} {120}},\ \bibinfo {pages}
  {051101} (\bibinfo {year} {2018})},\ \Eprint
  {http://arxiv.org/abs/1709.00219} {arXiv:1709.00219 [gr-qc]} \BibitemShut
  {NoStop}%
\bibitem [{\citenamefont {Tripathi}\ \emph {et~al.}(2018)\citenamefont
  {Tripathi}, \citenamefont {Nampalliwar}, \citenamefont {Abdikamalov},
  \citenamefont {Ayzenberg}, \citenamefont {Jiang},\ and\ \citenamefont
  {Bambi}}]{Tripathi2018a}%
  \BibitemOpen
  \bibfield  {author} {\bibinfo {author} {\bibfnamefont {A.}~\bibnamefont
  {Tripathi}}, \bibinfo {author} {\bibfnamefont {S.}~\bibnamefont
  {Nampalliwar}}, \bibinfo {author} {\bibfnamefont {A.~B.}\ \bibnamefont
  {Abdikamalov}}, \bibinfo {author} {\bibfnamefont {D.}~\bibnamefont
  {Ayzenberg}}, \bibinfo {author} {\bibfnamefont {J.}~\bibnamefont {Jiang}}, \
  and\ \bibinfo {author} {\bibfnamefont {C.}~\bibnamefont {Bambi}},\ }\href
  {\doibase 10.1103/PhysRevD.98.023018} {\bibfield  {journal} {\bibinfo
  {journal} {Phys. Rev.}\ }\textbf {\bibinfo {volume} {D98}},\ \bibinfo {pages}
  {023018} (\bibinfo {year} {2018})},\ \Eprint
  {http://arxiv.org/abs/1804.10380} {arXiv:1804.10380 [gr-qc]} \BibitemShut
  {NoStop}%
\bibitem [{\citenamefont {Xu}\ \emph {et~al.}(2018)\citenamefont {Xu},
  \citenamefont {Nampalliwar}, \citenamefont {Abdikamalov}, \citenamefont
  {Ayzenberg}, \citenamefont {Bambi}, \citenamefont {Dauser}, \citenamefont
  {Garcia},\ and\ \citenamefont {Jiang}}]{Xu2018}%
  \BibitemOpen
  \bibfield  {author} {\bibinfo {author} {\bibfnamefont {Y.}~\bibnamefont
  {Xu}}, \bibinfo {author} {\bibfnamefont {S.}~\bibnamefont {Nampalliwar}},
  \bibinfo {author} {\bibfnamefont {A.~B.}\ \bibnamefont {Abdikamalov}},
  \bibinfo {author} {\bibfnamefont {D.}~\bibnamefont {Ayzenberg}}, \bibinfo
  {author} {\bibfnamefont {C.}~\bibnamefont {Bambi}}, \bibinfo {author}
  {\bibfnamefont {T.}~\bibnamefont {Dauser}}, \bibinfo {author} {\bibfnamefont
  {J.~A.}\ \bibnamefont {Garcia}}, \ and\ \bibinfo {author} {\bibfnamefont
  {J.}~\bibnamefont {Jiang}},\ }\href {\doibase 10.3847/1538-4357/aadb9d}
  {\bibfield  {journal} {\bibinfo  {journal} {Astrophys. J.}\ }\textbf
  {\bibinfo {volume} {865}},\ \bibinfo {pages} {134} (\bibinfo {year}
  {2018})},\ \Eprint {http://arxiv.org/abs/1807.10243} {arXiv:1807.10243
  [gr-qc]} \BibitemShut {NoStop}%
\bibitem [{\citenamefont {Choudhury}\ \emph {et~al.}(2019)\citenamefont
  {Choudhury}, \citenamefont {Nampalliwar}, \citenamefont {Abdikamalov},
  \citenamefont {Ayzenberg}, \citenamefont {Bambi}, \citenamefont {Dauser},\
  and\ \citenamefont {Garcia}}]{Choudhury:2018zmf}%
  \BibitemOpen
  \bibfield  {author} {\bibinfo {author} {\bibfnamefont {K.}~\bibnamefont
  {Choudhury}}, \bibinfo {author} {\bibfnamefont {S.}~\bibnamefont
  {Nampalliwar}}, \bibinfo {author} {\bibfnamefont {A.~B.}\ \bibnamefont
  {Abdikamalov}}, \bibinfo {author} {\bibfnamefont {D.}~\bibnamefont
  {Ayzenberg}}, \bibinfo {author} {\bibfnamefont {C.}~\bibnamefont {Bambi}},
  \bibinfo {author} {\bibfnamefont {T.}~\bibnamefont {Dauser}}, \ and\ \bibinfo
  {author} {\bibfnamefont {J.~A.}\ \bibnamefont {Garcia}},\ }\href {\doibase
  10.3847/1538-4357/ab24d6} {\bibfield  {journal} {\bibinfo  {journal}
  {Astrophys. J.}\ }\textbf {\bibinfo {volume} {879}},\ \bibinfo {pages} {80}
  (\bibinfo {year} {2019})},\ \Eprint {http://arxiv.org/abs/1809.06669}
  {arXiv:1809.06669 [gr-qc]} \BibitemShut {NoStop}%
\bibitem [{\citenamefont {Zhou}\ \emph {et~al.}(2018)\citenamefont {Zhou},
  \citenamefont {Cao}, \citenamefont {Abdikamalov}, \citenamefont {Ayzenberg},
  \citenamefont {Bambi}, \citenamefont {Modesto},\ and\ \citenamefont
  {Nampalliwar}}]{Zhou2018a}%
  \BibitemOpen
  \bibfield  {author} {\bibinfo {author} {\bibfnamefont {M.}~\bibnamefont
  {Zhou}}, \bibinfo {author} {\bibfnamefont {Z.}~\bibnamefont {Cao}}, \bibinfo
  {author} {\bibfnamefont {A.}~\bibnamefont {Abdikamalov}}, \bibinfo {author}
  {\bibfnamefont {D.}~\bibnamefont {Ayzenberg}}, \bibinfo {author}
  {\bibfnamefont {C.}~\bibnamefont {Bambi}}, \bibinfo {author} {\bibfnamefont
  {L.}~\bibnamefont {Modesto}}, \ and\ \bibinfo {author} {\bibfnamefont
  {S.}~\bibnamefont {Nampalliwar}},\ }\href {\doibase
  10.1103/PhysRevD.98.024007} {\bibfield  {journal} {\bibinfo  {journal} {Phys.
  Rev.}\ }\textbf {\bibinfo {volume} {D98}},\ \bibinfo {pages} {024007}
  (\bibinfo {year} {2018})},\ \Eprint {http://arxiv.org/abs/1803.07849}
  {arXiv:1803.07849 [gr-qc]} \BibitemShut {NoStop}%
\bibitem [{\citenamefont {Zhou}\ \emph {et~al.}(2019)\citenamefont {Zhou},
  \citenamefont {Abdikamalov}, \citenamefont {Ayzenberg}, \citenamefont
  {Bambi}, \citenamefont {Modesto}, \citenamefont {Nampalliwar},\ and\
  \citenamefont {Xu}}]{Zhou2018b}%
  \BibitemOpen
  \bibfield  {author} {\bibinfo {author} {\bibfnamefont {M.}~\bibnamefont
  {Zhou}}, \bibinfo {author} {\bibfnamefont {A.}~\bibnamefont {Abdikamalov}},
  \bibinfo {author} {\bibfnamefont {D.}~\bibnamefont {Ayzenberg}}, \bibinfo
  {author} {\bibfnamefont {C.}~\bibnamefont {Bambi}}, \bibinfo {author}
  {\bibfnamefont {L.}~\bibnamefont {Modesto}}, \bibinfo {author} {\bibfnamefont
  {S.}~\bibnamefont {Nampalliwar}}, \ and\ \bibinfo {author} {\bibfnamefont
  {Y.}~\bibnamefont {Xu}},\ }\href {\doibase 10.1209/0295-5075/125/30002}
  {\bibfield  {journal} {\bibinfo  {journal} {EPL}\ }\textbf {\bibinfo {volume}
  {125}},\ \bibinfo {pages} {30002} (\bibinfo {year} {2019})}\BibitemShut
  {NoStop}%
\bibitem [{\citenamefont {Tripathi}\ \emph
  {et~al.}(2019{\natexlab{a}})\citenamefont {Tripathi}, \citenamefont
  {Nampalliwar}, \citenamefont {Abdikamalov}, \citenamefont {Ayzenberg},
  \citenamefont {Bambi}, \citenamefont {Dauser}, \citenamefont {Garcia},\ and\
  \citenamefont {Marinucci}}]{Tripathi:2018lhx}%
  \BibitemOpen
  \bibfield  {author} {\bibinfo {author} {\bibfnamefont {A.}~\bibnamefont
  {Tripathi}}, \bibinfo {author} {\bibfnamefont {S.}~\bibnamefont
  {Nampalliwar}}, \bibinfo {author} {\bibfnamefont {A.~B.}\ \bibnamefont
  {Abdikamalov}}, \bibinfo {author} {\bibfnamefont {D.}~\bibnamefont
  {Ayzenberg}}, \bibinfo {author} {\bibfnamefont {C.}~\bibnamefont {Bambi}},
  \bibinfo {author} {\bibfnamefont {T.}~\bibnamefont {Dauser}}, \bibinfo
  {author} {\bibfnamefont {J.~A.}\ \bibnamefont {Garcia}}, \ and\ \bibinfo
  {author} {\bibfnamefont {A.}~\bibnamefont {Marinucci}},\ }\href {\doibase
  10.3847/1538-4357/ab0e7e} {\bibfield  {journal} {\bibinfo  {journal}
  {Astrophys. J.}\ }\textbf {\bibinfo {volume} {875}},\ \bibinfo {pages} {56}
  (\bibinfo {year} {2019}{\natexlab{a}})},\ \Eprint
  {http://arxiv.org/abs/1811.08148} {arXiv:1811.08148 [gr-qc]} \BibitemShut
  {NoStop}%
\bibitem [{\citenamefont {Tripathi}\ \emph
  {et~al.}(2019{\natexlab{b}})\citenamefont {Tripathi} \emph
  {et~al.}}]{Tripathi:2019bya}%
  \BibitemOpen
  \bibfield  {author} {\bibinfo {author} {\bibfnamefont {A.}~\bibnamefont
  {Tripathi}} \emph {et~al.},\ }\href {\doibase 10.3847/1538-4357/ab0a00}
  {\bibfield  {journal} {\bibinfo  {journal} {Astrophys. J.}\ }\textbf
  {\bibinfo {volume} {874}},\ \bibinfo {pages} {135} (\bibinfo {year}
  {2019}{\natexlab{b}})},\ \Eprint {http://arxiv.org/abs/1901.03064}
  {arXiv:1901.03064 [gr-qc]} \BibitemShut {NoStop}%
\bibitem [{\citenamefont {Zhang}\ \emph
  {et~al.}(2019{\natexlab{a}})\citenamefont {Zhang}, \citenamefont
  {Abdikamalov}, \citenamefont {Ayzenberg}, \citenamefont {Bambi},
  \citenamefont {Dauser}, \citenamefont {Garcia},\ and\ \citenamefont
  {Nampalliwar}}]{Zhang:2019zsn}%
  \BibitemOpen
  \bibfield  {author} {\bibinfo {author} {\bibfnamefont {Y.}~\bibnamefont
  {Zhang}}, \bibinfo {author} {\bibfnamefont {A.~B.}\ \bibnamefont
  {Abdikamalov}}, \bibinfo {author} {\bibfnamefont {D.}~\bibnamefont
  {Ayzenberg}}, \bibinfo {author} {\bibfnamefont {C.}~\bibnamefont {Bambi}},
  \bibinfo {author} {\bibfnamefont {T.}~\bibnamefont {Dauser}}, \bibinfo
  {author} {\bibfnamefont {J.~A.}\ \bibnamefont {Garcia}}, \ and\ \bibinfo
  {author} {\bibfnamefont {S.}~\bibnamefont {Nampalliwar}},\ }\href {\doibase
  10.3847/1538-4357/ab0e79} {\bibfield  {journal} {\bibinfo  {journal}
  {Astrophys. J.}\ }\textbf {\bibinfo {volume} {875}},\ \bibinfo {pages} {41}
  (\bibinfo {year} {2019}{\natexlab{a}})},\ \Eprint
  {http://arxiv.org/abs/1901.06117} {arXiv:1901.06117 [gr-qc]} \BibitemShut
  {NoStop}%
\bibitem [{\citenamefont {Zhang}\ \emph
  {et~al.}(2019{\natexlab{b}})\citenamefont {Zhang}, \citenamefont
  {Abdikamalov}, \citenamefont {Ayzenberg}, \citenamefont {Bambi},\ and\
  \citenamefont {Nampalliwar}}]{Zhang:2019ldz}%
  \BibitemOpen
  \bibfield  {author} {\bibinfo {author} {\bibfnamefont {Y.}~\bibnamefont
  {Zhang}}, \bibinfo {author} {\bibfnamefont {A.~B.}\ \bibnamefont
  {Abdikamalov}}, \bibinfo {author} {\bibfnamefont {D.}~\bibnamefont
  {Ayzenberg}}, \bibinfo {author} {\bibfnamefont {C.}~\bibnamefont {Bambi}}, \
  and\ \bibinfo {author} {\bibfnamefont {S.}~\bibnamefont {Nampalliwar}},\
  }\href {\doibase 10.3847/1538-4357/ab4271} {\bibfield  {journal} {\bibinfo
  {journal} {Astrophys. J.}\ }\textbf {\bibinfo {volume} {884}},\ \bibinfo
  {pages} {147} (\bibinfo {year} {2019}{\natexlab{b}})},\ \Eprint
  {http://arxiv.org/abs/1907.03084} {arXiv:1907.03084 [gr-qc]} \BibitemShut
  {NoStop}%
\bibitem [{\citenamefont {Tripathi}\ \emph
  {et~al.}(2019{\natexlab{c}})\citenamefont {Tripathi}, \citenamefont
  {Abdikamalov}, \citenamefont {Ayzenberg}, \citenamefont {Bambi},\ and\
  \citenamefont {Nampalliwar}}]{Tripathi:2019fms}%
  \BibitemOpen
  \bibfield  {author} {\bibinfo {author} {\bibfnamefont {A.}~\bibnamefont
  {Tripathi}}, \bibinfo {author} {\bibfnamefont {A.~B.}\ \bibnamefont
  {Abdikamalov}}, \bibinfo {author} {\bibfnamefont {D.}~\bibnamefont
  {Ayzenberg}}, \bibinfo {author} {\bibfnamefont {C.}~\bibnamefont {Bambi}}, \
  and\ \bibinfo {author} {\bibfnamefont {S.}~\bibnamefont {Nampalliwar}},\
  }\href {\doibase 10.1103/PhysRevD.99.083001} {\bibfield  {journal} {\bibinfo
  {journal} {Phys. Rev. D}\ }\textbf {\bibinfo {volume} {99}},\ \bibinfo
  {pages} {083001} (\bibinfo {year} {2019}{\natexlab{c}})},\ \Eprint
  {http://arxiv.org/abs/1903.04071} {arXiv:1903.04071 [gr-qc]} \BibitemShut
  {NoStop}%
\bibitem [{rel({\natexlab{a}})}]{relxillnkweb}%
  \BibitemOpen
  \href@noop {} {\enquote {\bibinfo {title} {\textsc{relxill\_nk}},}\ }\bibinfo
  {howpublished}
  {\url{http://www.tat.physik.uni-tuebingen.de/~nampalliwar/relxill_nk/}}
  ({\natexlab{a}})\BibitemShut {NoStop}%
\bibitem [{rel({\natexlab{b}})}]{relxillnkweb2}%
  \BibitemOpen
  \href@noop {} {\enquote {\bibinfo {title} {\textsc{relxill\_nk}},}\ }\bibinfo
  {howpublished}
  {\url{http://www.physics.fudan.edu.cn/tps/people/bambi/Site/RELXILL_NK.html}}
  ({\natexlab{b}})\BibitemShut {NoStop}%
\bibitem [{\citenamefont {Bardeen}(1970)}]{Bardeen:1970zz}%
  \BibitemOpen
  \bibfield  {author} {\bibinfo {author} {\bibfnamefont {J.~M.}\ \bibnamefont
  {Bardeen}},\ }\href {\doibase 10.1038/226064a0} {\bibfield  {journal}
  {\bibinfo  {journal} {Nature}\ }\textbf {\bibinfo {volume} {226}},\ \bibinfo
  {pages} {64} (\bibinfo {year} {1970})}\BibitemShut {NoStop}%
\bibitem [{\citenamefont {Thorne}(1974)}]{Thorne:1974ve}%
  \BibitemOpen
  \bibfield  {author} {\bibinfo {author} {\bibfnamefont {K.~S.}\ \bibnamefont
  {Thorne}},\ }\href {\doibase 10.1086/152991} {\bibfield  {journal} {\bibinfo
  {journal} {Astrophys. J.}\ }\textbf {\bibinfo {volume} {191}},\ \bibinfo
  {pages} {507} (\bibinfo {year} {1974})}\BibitemShut {NoStop}%
\bibitem [{\citenamefont {Gammie}\ \emph {et~al.}(2004)\citenamefont {Gammie},
  \citenamefont {Shapiro},\ and\ \citenamefont {McKinney}}]{Gammie:2003qi}%
  \BibitemOpen
  \bibfield  {author} {\bibinfo {author} {\bibfnamefont {C.~F.}\ \bibnamefont
  {Gammie}}, \bibinfo {author} {\bibfnamefont {S.~L.}\ \bibnamefont {Shapiro}},
  \ and\ \bibinfo {author} {\bibfnamefont {J.~C.}\ \bibnamefont {McKinney}},\
  }\href {\doibase 10.1086/380996} {\bibfield  {journal} {\bibinfo  {journal}
  {Astrophys. J.}\ }\textbf {\bibinfo {volume} {602}},\ \bibinfo {pages} {312}
  (\bibinfo {year} {2004})},\ \Eprint {http://arxiv.org/abs/astro-ph/0310886}
  {arXiv:astro-ph/0310886} \BibitemShut {NoStop}%
\bibitem [{\citenamefont {Dauser}\ \emph {et~al.}(2013)\citenamefont {Dauser},
  \citenamefont {Garcia}, \citenamefont {Wilms}, \citenamefont {Bock},
  \citenamefont {Brenneman}, \citenamefont {Falanga}, \citenamefont
  {Fukumura},\ and\ \citenamefont {Reynolds}}]{Dauser:2013xv}%
  \BibitemOpen
  \bibfield  {author} {\bibinfo {author} {\bibfnamefont {T.}~\bibnamefont
  {Dauser}}, \bibinfo {author} {\bibfnamefont {J.}~\bibnamefont {Garcia}},
  \bibinfo {author} {\bibfnamefont {J.}~\bibnamefont {Wilms}}, \bibinfo
  {author} {\bibfnamefont {M.}~\bibnamefont {Bock}}, \bibinfo {author}
  {\bibfnamefont {L.~W.}\ \bibnamefont {Brenneman}}, \bibinfo {author}
  {\bibfnamefont {M.}~\bibnamefont {Falanga}}, \bibinfo {author} {\bibfnamefont
  {K.}~\bibnamefont {Fukumura}}, \ and\ \bibinfo {author} {\bibfnamefont
  {C.~S.}\ \bibnamefont {Reynolds}},\ }\href {\doibase 10.1093/mnras/sts710}
  {\bibfield  {journal} {\bibinfo  {journal} {Mon. Not. Roy. Astron. Soc.}\
  }\textbf {\bibinfo {volume} {430}},\ \bibinfo {pages} {1694} (\bibinfo {year}
  {2013})},\ \Eprint {http://arxiv.org/abs/1301.4922} {arXiv:1301.4922
  [astro-ph.HE]} \BibitemShut {NoStop}%
\bibitem [{\citenamefont {Nampalliwar}\ and\ \citenamefont
  {Bambi}(2018)}]{Bambi:2018thh}%
  \BibitemOpen
  \bibfield  {author} {\bibinfo {author} {\bibfnamefont {S.}~\bibnamefont
  {Nampalliwar}}\ and\ \bibinfo {author} {\bibfnamefont {C.}~\bibnamefont
  {Bambi}},\ }\href@noop {} {\  (\bibinfo {year} {2018})},\ \Eprint
  {http://arxiv.org/abs/1810.07041} {arXiv:1810.07041 [astro-ph.HE]}
  \BibitemShut {NoStop}%
\bibitem [{\citenamefont {Novikov}\ and\ \citenamefont
  {Thorne}(1973)}]{Novikov1973}%
  \BibitemOpen
  \bibfield  {author} {\bibinfo {author} {\bibfnamefont {I.~D.}\ \bibnamefont
  {Novikov}}\ and\ \bibinfo {author} {\bibfnamefont {K.~S.}\ \bibnamefont
  {Thorne}},\ }in\ \href@noop {} {\emph {\bibinfo {booktitle} {{Proceedings,
  Ecole d'Et\'{e} de Physique Th\'{e}orique: Les Astres Occlus: Les Houches,
  France, August, 1972}}}}\ (\bibinfo {year} {1973})\ pp.\ \bibinfo {pages}
  {343--550}\BibitemShut {NoStop}%
\bibitem [{\citenamefont {Wilkins}\ and\ \citenamefont
  {Fabian}(2012)}]{Wilkins2012}%
  \BibitemOpen
  \bibfield  {author} {\bibinfo {author} {\bibfnamefont {D.~R.}\ \bibnamefont
  {Wilkins}}\ and\ \bibinfo {author} {\bibfnamefont {A.~C.}\ \bibnamefont
  {Fabian}},\ }\href {\doibase 10.1111/j.1365-2966.2012.21308.x} {\bibfield
  {journal} {\bibinfo  {journal} {Mon. Not. Roy. Astron. Soc.}\ }\textbf
  {\bibinfo {volume} {424}},\ \bibinfo {pages} {1284} (\bibinfo {year}
  {2012})},\ \Eprint {http://arxiv.org/abs/1205.3179} {arXiv:1205.3179
  [astro-ph.HE]} \BibitemShut {NoStop}%
\bibitem [{\citenamefont {Wilkins}\ and\ \citenamefont
  {Gallo}(2015)}]{Wilkins2014}%
  \BibitemOpen
  \bibfield  {author} {\bibinfo {author} {\bibfnamefont {D.~R.}\ \bibnamefont
  {Wilkins}}\ and\ \bibinfo {author} {\bibfnamefont {L.~C.}\ \bibnamefont
  {Gallo}},\ }\href {\doibase 10.1093/mnras/stu2524} {\bibfield  {journal}
  {\bibinfo  {journal} {Mon. Not. Roy. Astron. Soc.}\ }\textbf {\bibinfo
  {volume} {448}},\ \bibinfo {pages} {703} (\bibinfo {year} {2015})},\ \Eprint
  {http://arxiv.org/abs/1412.0015} {arXiv:1412.0015 [astro-ph.HE]} \BibitemShut
  {NoStop}%
\bibitem [{\citenamefont {Kara}\ \emph {et~al.}(2016)\citenamefont {Kara},
  \citenamefont {Miller}, \citenamefont {Reynolds},\ and\ \citenamefont
  {Dai}}]{Kara:2016kbu}%
  \BibitemOpen
  \bibfield  {author} {\bibinfo {author} {\bibfnamefont {E.}~\bibnamefont
  {Kara}}, \bibinfo {author} {\bibfnamefont {J.~M.}\ \bibnamefont {Miller}},
  \bibinfo {author} {\bibfnamefont {C.}~\bibnamefont {Reynolds}}, \ and\
  \bibinfo {author} {\bibfnamefont {L.}~\bibnamefont {Dai}},\ }\href {\doibase
  10.1038/nature18007} {\bibfield  {journal} {\bibinfo  {journal} {Nature}\
  }\textbf {\bibinfo {volume} {535}},\ \bibinfo {pages} {388} (\bibinfo {year}
  {2016})},\ \Eprint {http://arxiv.org/abs/1606.06736} {arXiv:1606.06736
  [astro-ph.HE]} \BibitemShut {NoStop}%
\bibitem [{\citenamefont {Wilkins}\ and\ \citenamefont
  {Fabian}(2011)}]{Wilkins:2011kt}%
  \BibitemOpen
  \bibfield  {author} {\bibinfo {author} {\bibfnamefont {D.}~\bibnamefont
  {Wilkins}}\ and\ \bibinfo {author} {\bibfnamefont {A.}~\bibnamefont
  {Fabian}},\ }\href {\doibase 10.1111/j.1365-2966.2011.18458.x} {\bibfield
  {journal} {\bibinfo  {journal} {Mon. Not. Roy. Astron. Soc.}\ }\textbf
  {\bibinfo {volume} {414}},\ \bibinfo {pages} {1269} (\bibinfo {year}
  {2011})},\ \Eprint {http://arxiv.org/abs/1102.0433} {arXiv:1102.0433
  [astro-ph.HE]} \BibitemShut {NoStop}%
\bibitem [{\citenamefont {Bambi}(2012)}]{Bambi:2012tg}%
  \BibitemOpen
  \bibfield  {author} {\bibinfo {author} {\bibfnamefont {C.}~\bibnamefont
  {Bambi}},\ }\href {\doibase 10.1088/0004-637X/761/2/174} {\bibfield
  {journal} {\bibinfo  {journal} {Astrophys. J.}\ }\textbf {\bibinfo {volume}
  {761}},\ \bibinfo {pages} {174} (\bibinfo {year} {2012})},\ \Eprint
  {http://arxiv.org/abs/1210.5679} {arXiv:1210.5679 [gr-qc]} \BibitemShut
  {NoStop}%
\bibitem [{\citenamefont {{Speith}}\ \emph {et~al.}(1995)\citenamefont
  {{Speith}}, \citenamefont {{Riffert}},\ and\ \citenamefont
  {{Ruder}}}]{speith1995}%
  \BibitemOpen
  \bibfield  {author} {\bibinfo {author} {\bibfnamefont {R.}~\bibnamefont
  {{Speith}}}, \bibinfo {author} {\bibfnamefont {H.}~\bibnamefont {{Riffert}}},
  \ and\ \bibinfo {author} {\bibfnamefont {H.}~\bibnamefont {{Ruder}}},\ }\href
  {\doibase 10.1016/0010-4655(95)00067-P} {\bibfield  {journal} {\bibinfo
  {journal} {Computer Physics Communications}\ }\textbf {\bibinfo {volume}
  {88}},\ \bibinfo {pages} {109} (\bibinfo {year} {1995})}\BibitemShut
  {NoStop}%
\bibitem [{\citenamefont {Cunningham}(1975)}]{Cunningham1975}%
  \BibitemOpen
  \bibfield  {author} {\bibinfo {author} {\bibfnamefont {C.~T.}\ \bibnamefont
  {Cunningham}},\ }\href {\doibase 10.1086/154033} {\bibfield  {journal}
  {\bibinfo  {journal} {Astrophys. J.}\ }\textbf {\bibinfo {volume} {202}},\
  \bibinfo {pages} {788} (\bibinfo {year} {1975})}\BibitemShut {NoStop}%
\bibitem [{\citenamefont {Reynolds}\ and\ \citenamefont
  {Fabian}(2008)}]{Reynolds:2007rx}%
  \BibitemOpen
  \bibfield  {author} {\bibinfo {author} {\bibfnamefont {C.~S.}\ \bibnamefont
  {Reynolds}}\ and\ \bibinfo {author} {\bibfnamefont {A.~C.}\ \bibnamefont
  {Fabian}},\ }\href {\doibase 10.1086/527344} {\bibfield  {journal} {\bibinfo
  {journal} {Astrophys. J.}\ }\textbf {\bibinfo {volume} {675}},\ \bibinfo
  {pages} {1048} (\bibinfo {year} {2008})},\ \Eprint
  {http://arxiv.org/abs/0711.4158} {arXiv:0711.4158 [astro-ph]} \BibitemShut
  {NoStop}%
\bibitem [{\citenamefont {Penna}\ \emph {et~al.}(2010)\citenamefont {Penna}
  \emph {et~al.}}]{Penna:2010hu}%
  \BibitemOpen
  \bibfield  {author} {\bibinfo {author} {\bibfnamefont {R.~F.}\ \bibnamefont
  {Penna}} \emph {et~al.},\ }\href {\doibase 10.1111/j.1365-2966.2010.17170.x}
  {\bibfield  {journal} {\bibinfo  {journal} {MNRAS}\ }\textbf {\bibinfo
  {volume} {408}},\ \bibinfo {pages} {752} (\bibinfo {year} {2010})},\ \Eprint
  {http://arxiv.org/abs/1003.0966} {arXiv:1003.0966 [astro-ph.HE]} \BibitemShut
  {NoStop}%
\bibitem [{\citenamefont {Steiner}\ \emph {et~al.}(2010)\citenamefont
  {Steiner}, \citenamefont {McClintock}, \citenamefont {Remillard},
  \citenamefont {Gou}, \citenamefont {Yamada},\ and\ \citenamefont
  {Narayan}}]{Steiner:2010kd}%
  \BibitemOpen
  \bibfield  {author} {\bibinfo {author} {\bibfnamefont {J.~F.}\ \bibnamefont
  {Steiner}}, \bibinfo {author} {\bibfnamefont {J.~E.}\ \bibnamefont
  {McClintock}}, \bibinfo {author} {\bibfnamefont {R.~A.}\ \bibnamefont
  {Remillard}}, \bibinfo {author} {\bibfnamefont {L.}~\bibnamefont {Gou}},
  \bibinfo {author} {\bibfnamefont {S.}~\bibnamefont {Yamada}}, \ and\ \bibinfo
  {author} {\bibfnamefont {R.}~\bibnamefont {Narayan}},\ }\href {\doibase
  10.1088/2041-8205/718/2/L117} {\bibfield  {journal} {\bibinfo  {journal}
  {Astrophys. J.}\ }\textbf {\bibinfo {volume} {718}},\ \bibinfo {pages} {L117}
  (\bibinfo {year} {2010})},\ \Eprint {http://arxiv.org/abs/1006.5729}
  {arXiv:1006.5729 [astro-ph.HE]} \BibitemShut {NoStop}%
\bibitem [{\citenamefont {Done}\ \emph {et~al.}(2007)\citenamefont {Done},
  \citenamefont {Gierlinski},\ and\ \citenamefont {Kubota}}]{Done:2007nc}%
  \BibitemOpen
  \bibfield  {author} {\bibinfo {author} {\bibfnamefont {C.}~\bibnamefont
  {Done}}, \bibinfo {author} {\bibfnamefont {M.}~\bibnamefont {Gierlinski}}, \
  and\ \bibinfo {author} {\bibfnamefont {A.}~\bibnamefont {Kubota}},\ }\href
  {\doibase 10.1007/s00159-007-0006-1} {\bibfield  {journal} {\bibinfo
  {journal} {Astron. Astrophys. Rev.}\ }\textbf {\bibinfo {volume} {15}},\
  \bibinfo {pages} {1} (\bibinfo {year} {2007})},\ \Eprint
  {http://arxiv.org/abs/0708.0148} {arXiv:0708.0148 [astro-ph]} \BibitemShut
  {NoStop}%
\bibitem [{\citenamefont {Zdziarski}\ and\ \citenamefont
  {De~Marco}(2020)}]{Zdziarski:2020pgi}%
  \BibitemOpen
  \bibfield  {author} {\bibinfo {author} {\bibfnamefont {A.~A.}\ \bibnamefont
  {Zdziarski}}\ and\ \bibinfo {author} {\bibfnamefont {B.}~\bibnamefont
  {De~Marco}},\ }\href {\doibase 10.3847/2041-8213/ab9899} {\bibfield
  {journal} {\bibinfo  {journal} {Astrophys. J. Lett.}\ }\textbf {\bibinfo
  {volume} {896}},\ \bibinfo {pages} {L36} (\bibinfo {year} {2020})},\ \Eprint
  {http://arxiv.org/abs/2002.04652} {arXiv:2002.04652 [astro-ph.HE]}
  \BibitemShut {NoStop}%
\bibitem [{\citenamefont {Wilkins}\ \emph {et~al.}(2020)\citenamefont
  {Wilkins}, \citenamefont {Reynolds},\ and\ \citenamefont
  {Fabian}}]{Wilkins:2020pgu}%
  \BibitemOpen
  \bibfield  {author} {\bibinfo {author} {\bibfnamefont {D.}~\bibnamefont
  {Wilkins}}, \bibinfo {author} {\bibfnamefont {C.}~\bibnamefont {Reynolds}}, \
  and\ \bibinfo {author} {\bibfnamefont {A.}~\bibnamefont {Fabian}},\ }\href
  {\doibase 10.1093/mnras/staa628} {\bibfield  {journal} {\bibinfo  {journal}
  {Mon. Not. Roy. Astron. Soc.}\ }\textbf {\bibinfo {volume} {493}},\ \bibinfo
  {pages} {5532} (\bibinfo {year} {2020})},\ \Eprint
  {http://arxiv.org/abs/2003.00019} {arXiv:2003.00019 [astro-ph.HE]}
  \BibitemShut {NoStop}%
\bibitem [{\citenamefont {Garc\'\i{}a}\ \emph {et~al.}(2018)\citenamefont
  {Garc\'\i{}a}, \citenamefont {Kallman}, \citenamefont {Bautista},
  \citenamefont {Mendoza}, \citenamefont {Deprince}, \citenamefont {Palmeri},\
  and\ \citenamefont {Quinet}}]{Garcia:2018czl}%
  \BibitemOpen
  \bibfield  {author} {\bibinfo {author} {\bibfnamefont {J.}~\bibnamefont
  {Garc\'\i{}a}}, \bibinfo {author} {\bibfnamefont {T.}~\bibnamefont
  {Kallman}}, \bibinfo {author} {\bibfnamefont {M.}~\bibnamefont {Bautista}},
  \bibinfo {author} {\bibfnamefont {C.}~\bibnamefont {Mendoza}}, \bibinfo
  {author} {\bibfnamefont {J.}~\bibnamefont {Deprince}}, \bibinfo {author}
  {\bibfnamefont {P.}~\bibnamefont {Palmeri}}, \ and\ \bibinfo {author}
  {\bibfnamefont {P.}~\bibnamefont {Quinet}},\ }\href@noop {} {\bibfield
  {journal} {\bibinfo  {journal} {ASP Conf. Ser.}\ }\textbf {\bibinfo {volume}
  {515}},\ \bibinfo {pages} {282} (\bibinfo {year} {2018})},\ \Eprint
  {http://arxiv.org/abs/1805.00581} {arXiv:1805.00581 [astro-ph.HE]}
  \BibitemShut {NoStop}%
\bibitem [{\citenamefont {Reynolds}\ \emph {et~al.}(2012)\citenamefont
  {Reynolds}, \citenamefont {Brenneman}, \citenamefont {Lohfink}, \citenamefont
  {Trippe}, \citenamefont {Miller}, \citenamefont {Fabian},\ and\ \citenamefont
  {Nowak}}]{Reynolds:2012ji}%
  \BibitemOpen
  \bibfield  {author} {\bibinfo {author} {\bibfnamefont {C.}~\bibnamefont
  {Reynolds}}, \bibinfo {author} {\bibfnamefont {L.}~\bibnamefont {Brenneman}},
  \bibinfo {author} {\bibfnamefont {A.}~\bibnamefont {Lohfink}}, \bibinfo
  {author} {\bibfnamefont {M.}~\bibnamefont {Trippe}}, \bibinfo {author}
  {\bibfnamefont {J.}~\bibnamefont {Miller}}, \bibinfo {author} {\bibfnamefont
  {A.}~\bibnamefont {Fabian}}, \ and\ \bibinfo {author} {\bibfnamefont
  {M.}~\bibnamefont {Nowak}},\ }\href {\doibase 10.1088/0004-637X/755/2/88}
  {\bibfield  {journal} {\bibinfo  {journal} {Astrophys. J.}\ }\textbf
  {\bibinfo {volume} {755}},\ \bibinfo {pages} {88} (\bibinfo {year} {2012})},\
  \Eprint {http://arxiv.org/abs/1204.5747} {arXiv:1204.5747 [astro-ph.HE]}
  \BibitemShut {NoStop}%
\bibitem [{\citenamefont {Fabian}\ \emph {et~al.}(2002)\citenamefont {Fabian},
  \citenamefont {Vaughan}, \citenamefont {Nandra}, \citenamefont {Iwasawa},
  \citenamefont {Ballantyne}, \citenamefont {Lee}, \citenamefont {De~Rosa},
  \citenamefont {Turner},\ and\ \citenamefont {Young}}]{Fabian:2002gj}%
  \BibitemOpen
  \bibfield  {author} {\bibinfo {author} {\bibfnamefont {A.}~\bibnamefont
  {Fabian}}, \bibinfo {author} {\bibfnamefont {S.}~\bibnamefont {Vaughan}},
  \bibinfo {author} {\bibfnamefont {K.}~\bibnamefont {Nandra}}, \bibinfo
  {author} {\bibfnamefont {K.}~\bibnamefont {Iwasawa}}, \bibinfo {author}
  {\bibfnamefont {D.}~\bibnamefont {Ballantyne}}, \bibinfo {author}
  {\bibfnamefont {J.}~\bibnamefont {Lee}}, \bibinfo {author} {\bibfnamefont
  {A.}~\bibnamefont {De~Rosa}}, \bibinfo {author} {\bibfnamefont
  {A.}~\bibnamefont {Turner}}, \ and\ \bibinfo {author} {\bibfnamefont
  {A.}~\bibnamefont {Young}},\ }\href {\doibase
  10.1046/j.1365-8711.2002.05740.x} {\bibfield  {journal} {\bibinfo  {journal}
  {Mon. Not. Roy. Astron. Soc.}\ }\textbf {\bibinfo {volume} {335}},\ \bibinfo
  {pages} {L1} (\bibinfo {year} {2002})},\ \Eprint
  {http://arxiv.org/abs/astro-ph/0206095} {arXiv:astro-ph/0206095} \BibitemShut
  {NoStop}%
\bibitem [{\citenamefont {Waddell}\ and\ \citenamefont
  {Gallo}(2020)}]{Waddell:2020dqr}%
  \BibitemOpen
  \bibfield  {author} {\bibinfo {author} {\bibfnamefont {S.~G.}\ \bibnamefont
  {Waddell}}\ and\ \bibinfo {author} {\bibfnamefont {L.~C.}\ \bibnamefont
  {Gallo}},\ }\href@noop {} {\  (\bibinfo {year} {2020})},\ \Eprint
  {http://arxiv.org/abs/2009.04378} {arXiv:2009.04378 [astro-ph.HE]}
  \BibitemShut {NoStop}%
\bibitem [{\citenamefont {Reid}\ \emph {et~al.}(2014)\citenamefont {Reid},
  \citenamefont {McClintock}, \citenamefont {Steiner}, \citenamefont {Steeghs},
  \citenamefont {Remillard}, \citenamefont {Dhawan},\ and\ \citenamefont
  {Narayan}}]{Reid:2014ywa}%
  \BibitemOpen
  \bibfield  {author} {\bibinfo {author} {\bibfnamefont {M.}~\bibnamefont
  {Reid}}, \bibinfo {author} {\bibfnamefont {J.}~\bibnamefont {McClintock}},
  \bibinfo {author} {\bibfnamefont {J.}~\bibnamefont {Steiner}}, \bibinfo
  {author} {\bibfnamefont {D.}~\bibnamefont {Steeghs}}, \bibinfo {author}
  {\bibfnamefont {R.}~\bibnamefont {Remillard}}, \bibinfo {author}
  {\bibfnamefont {V.}~\bibnamefont {Dhawan}}, \ and\ \bibinfo {author}
  {\bibfnamefont {R.}~\bibnamefont {Narayan}},\ }\href {\doibase
  10.1088/0004-637X/796/1/2} {\bibfield  {journal} {\bibinfo  {journal}
  {Astrophys. J.}\ }\textbf {\bibinfo {volume} {796}},\ \bibinfo {pages} {2}
  (\bibinfo {year} {2014})},\ \Eprint {http://arxiv.org/abs/1409.2453}
  {arXiv:1409.2453 [astro-ph.GA]} \BibitemShut {NoStop}%
\bibitem [{\citenamefont {Miniutti}\ \emph {et~al.}(2003)\citenamefont
  {Miniutti}, \citenamefont {Fabian}, \citenamefont {Goyder},\ and\
  \citenamefont {Lasenby}}]{Miniutti:2003yd}%
  \BibitemOpen
  \bibfield  {author} {\bibinfo {author} {\bibfnamefont {G.}~\bibnamefont
  {Miniutti}}, \bibinfo {author} {\bibfnamefont {A.}~\bibnamefont {Fabian}},
  \bibinfo {author} {\bibfnamefont {R.}~\bibnamefont {Goyder}}, \ and\ \bibinfo
  {author} {\bibfnamefont {A.}~\bibnamefont {Lasenby}},\ }\href {\doibase
  10.1046/j.1365-8711.2003.06988.x} {\bibfield  {journal} {\bibinfo  {journal}
  {Mon. Not. Roy. Astron. Soc.}\ }\textbf {\bibinfo {volume} {344}},\ \bibinfo
  {pages} {L22} (\bibinfo {year} {2003})},\ \Eprint
  {http://arxiv.org/abs/astro-ph/0307163} {arXiv:astro-ph/0307163} \BibitemShut
  {NoStop}%
\bibitem [{\citenamefont {Zhu}\ \emph {et~al.}(2020)\citenamefont {Zhu},
  \citenamefont {Abdikamalov}, \citenamefont {Ayzenberg}, \citenamefont
  {Azreg-Ainou}, \citenamefont {Bambi}, \citenamefont {Jamil}, \citenamefont
  {Nampalliwar}, \citenamefont {Tripathi},\ and\ \citenamefont
  {Zhou}}]{Zhu:2020cfn}%
  \BibitemOpen
  \bibfield  {author} {\bibinfo {author} {\bibfnamefont {J.}~\bibnamefont
  {Zhu}}, \bibinfo {author} {\bibfnamefont {A.~B.}\ \bibnamefont
  {Abdikamalov}}, \bibinfo {author} {\bibfnamefont {D.}~\bibnamefont
  {Ayzenberg}}, \bibinfo {author} {\bibfnamefont {M.}~\bibnamefont
  {Azreg-Ainou}}, \bibinfo {author} {\bibfnamefont {C.}~\bibnamefont {Bambi}},
  \bibinfo {author} {\bibfnamefont {M.}~\bibnamefont {Jamil}}, \bibinfo
  {author} {\bibfnamefont {S.}~\bibnamefont {Nampalliwar}}, \bibinfo {author}
  {\bibfnamefont {A.}~\bibnamefont {Tripathi}}, \ and\ \bibinfo {author}
  {\bibfnamefont {M.}~\bibnamefont {Zhou}},\ }\href {\doibase
  10.1140/epjc/s10052-020-8198-x} {\bibfield  {journal} {\bibinfo  {journal}
  {Eur. Phys. J. C}\ }\textbf {\bibinfo {volume} {80}},\ \bibinfo {pages} {622}
  (\bibinfo {year} {2020})},\ \Eprint {http://arxiv.org/abs/2005.00184}
  {arXiv:2005.00184 [gr-qc]} \BibitemShut {NoStop}%
\bibitem [{\citenamefont {Abdikamalov}\ \emph {et~al.}(2020)\citenamefont
  {Abdikamalov}, \citenamefont {Ayzenberg}, \citenamefont {Bambi},
  \citenamefont {Dauser}, \citenamefont {Garcia}, \citenamefont {Nampalliwar},
  \citenamefont {Tripathi},\ and\ \citenamefont {Zhou}}]{Abdikamalov:2020oci}%
  \BibitemOpen
  \bibfield  {author} {\bibinfo {author} {\bibfnamefont {A.~B.}\ \bibnamefont
  {Abdikamalov}}, \bibinfo {author} {\bibfnamefont {D.}~\bibnamefont
  {Ayzenberg}}, \bibinfo {author} {\bibfnamefont {C.}~\bibnamefont {Bambi}},
  \bibinfo {author} {\bibfnamefont {T.}~\bibnamefont {Dauser}}, \bibinfo
  {author} {\bibfnamefont {J.~A.}\ \bibnamefont {Garcia}}, \bibinfo {author}
  {\bibfnamefont {S.}~\bibnamefont {Nampalliwar}}, \bibinfo {author}
  {\bibfnamefont {A.}~\bibnamefont {Tripathi}}, \ and\ \bibinfo {author}
  {\bibfnamefont {M.}~\bibnamefont {Zhou}},\ }\href {\doibase
  10.3847/1538-4357/aba625} {\bibfield  {journal} {\bibinfo  {journal}
  {Astrophys. J.}\ }\textbf {\bibinfo {volume} {899}},\ \bibinfo {pages} {80}
  (\bibinfo {year} {2020})},\ \Eprint {http://arxiv.org/abs/2003.09663}
  {arXiv:2003.09663 [astro-ph.HE]} \BibitemShut {NoStop}%
\bibitem [{\citenamefont {Tripathi}\ \emph {et~al.}(2020)\citenamefont
  {Tripathi}, \citenamefont {Zhou}, \citenamefont {Abdikamalov}, \citenamefont
  {Ayzenberg}, \citenamefont {Bambi},\ and\ \citenamefont
  {Nampalliwar}}]{Tripathi:2020wfi}%
  \BibitemOpen
  \bibfield  {author} {\bibinfo {author} {\bibfnamefont {A.}~\bibnamefont
  {Tripathi}}, \bibinfo {author} {\bibfnamefont {B.}~\bibnamefont {Zhou}},
  \bibinfo {author} {\bibfnamefont {A.~B.}\ \bibnamefont {Abdikamalov}},
  \bibinfo {author} {\bibfnamefont {D.}~\bibnamefont {Ayzenberg}}, \bibinfo
  {author} {\bibfnamefont {C.}~\bibnamefont {Bambi}}, \ and\ \bibinfo {author}
  {\bibfnamefont {S.}~\bibnamefont {Nampalliwar}},\ }\href@noop {} {\
  (\bibinfo {year} {2020})},\ \Eprint {http://arxiv.org/abs/2008.01934}
  {arXiv:2008.01934 [astro-ph.HE]} \BibitemShut {NoStop}%
\bibitem [{\citenamefont {Blum}\ \emph {et~al.}(2009)\citenamefont {Blum},
  \citenamefont {Miller}, \citenamefont {Fabian}, \citenamefont {Miller},
  \citenamefont {Homan}, \citenamefont {van~der Klis}, \citenamefont
  {Cackett},\ and\ \citenamefont {Reis}}]{Blum:2009ez}%
  \BibitemOpen
  \bibfield  {author} {\bibinfo {author} {\bibfnamefont {J.}~\bibnamefont
  {Blum}}, \bibinfo {author} {\bibfnamefont {J.}~\bibnamefont {Miller}},
  \bibinfo {author} {\bibfnamefont {A.}~\bibnamefont {Fabian}}, \bibinfo
  {author} {\bibfnamefont {M.}~\bibnamefont {Miller}}, \bibinfo {author}
  {\bibfnamefont {J.}~\bibnamefont {Homan}}, \bibinfo {author} {\bibfnamefont
  {M.}~\bibnamefont {van~der Klis}}, \bibinfo {author} {\bibfnamefont
  {E.}~\bibnamefont {Cackett}}, \ and\ \bibinfo {author} {\bibfnamefont
  {R.}~\bibnamefont {Reis}},\ }\href {\doibase 10.1088/0004-637X/706/1/60}
  {\bibfield  {journal} {\bibinfo  {journal} {Astrophys. J.}\ }\textbf
  {\bibinfo {volume} {706}},\ \bibinfo {pages} {60} (\bibinfo {year} {2009})},\
  \Eprint {http://arxiv.org/abs/0909.5383} {arXiv:0909.5383 [astro-ph.HE]}
  \BibitemShut {NoStop}%
\bibitem [{\citenamefont {Wilms}\ \emph {et~al.}(2000)\citenamefont {Wilms},
  \citenamefont {Allen},\ and\ \citenamefont {McCray}}]{Wilms2000}%
  \BibitemOpen
  \bibfield  {author} {\bibinfo {author} {\bibfnamefont {J.}~\bibnamefont
  {Wilms}}, \bibinfo {author} {\bibfnamefont {A.}~\bibnamefont {Allen}}, \ and\
  \bibinfo {author} {\bibfnamefont {R.}~\bibnamefont {McCray}},\ }\href
  {\doibase 10.1086/317016} {\bibfield  {journal} {\bibinfo  {journal}
  {Astrophys. J.}\ }\textbf {\bibinfo {volume} {542}},\ \bibinfo {pages} {914}
  (\bibinfo {year} {2000})},\ \Eprint {http://arxiv.org/abs/astro-ph/0008425}
  {arXiv:astro-ph/0008425 [astro-ph]} \BibitemShut {NoStop}%
\bibitem [{\citenamefont {Kulkarni}\ \emph {et~al.}(2011)\citenamefont
  {Kulkarni}, \citenamefont {Penna}, \citenamefont {Shcherbakov}, \citenamefont
  {Steiner}, \citenamefont {Narayan}, \citenamefont {Sadowski}, \citenamefont
  {Zhu}, \citenamefont {McClintock}, \citenamefont {Davis},\ and\ \citenamefont
  {McKinney}}]{Kulkarni:2011cy}%
  \BibitemOpen
  \bibfield  {author} {\bibinfo {author} {\bibfnamefont {A.~K.}\ \bibnamefont
  {Kulkarni}}, \bibinfo {author} {\bibfnamefont {R.~F.}\ \bibnamefont {Penna}},
  \bibinfo {author} {\bibfnamefont {R.~V.}\ \bibnamefont {Shcherbakov}},
  \bibinfo {author} {\bibfnamefont {J.~F.}\ \bibnamefont {Steiner}}, \bibinfo
  {author} {\bibfnamefont {R.}~\bibnamefont {Narayan}}, \bibinfo {author}
  {\bibfnamefont {A.}~\bibnamefont {Sadowski}}, \bibinfo {author}
  {\bibfnamefont {Y.}~\bibnamefont {Zhu}}, \bibinfo {author} {\bibfnamefont
  {J.~E.}\ \bibnamefont {McClintock}}, \bibinfo {author} {\bibfnamefont
  {S.~W.}\ \bibnamefont {Davis}}, \ and\ \bibinfo {author} {\bibfnamefont
  {J.~C.}\ \bibnamefont {McKinney}},\ }\href {\doibase
  10.1111/j.1365-2966.2011.18446.x} {\bibfield  {journal} {\bibinfo  {journal}
  {Mon. Not. Roy. Astron. Soc.}\ }\textbf {\bibinfo {volume} {414}},\ \bibinfo
  {pages} {1183} (\bibinfo {year} {2011})},\ \Eprint
  {http://arxiv.org/abs/1102.0010} {arXiv:1102.0010 [astro-ph.HE]} \BibitemShut
  {NoStop}%
\bibitem [{\citenamefont {Miller}\ \emph {et~al.}(2013)\citenamefont {Miller}
  \emph {et~al.}}]{Miller:2013rca}%
  \BibitemOpen
  \bibfield  {author} {\bibinfo {author} {\bibfnamefont {J.}~\bibnamefont
  {Miller}} \emph {et~al.},\ }\href {\doibase 10.1088/2041-8205/775/2/L45}
  {\bibfield  {journal} {\bibinfo  {journal} {Astrophys. J. Lett.}\ }\textbf
  {\bibinfo {volume} {775}},\ \bibinfo {pages} {L45} (\bibinfo {year}
  {2013})},\ \Eprint {http://arxiv.org/abs/1308.4669} {arXiv:1308.4669
  [astro-ph.HE]} \BibitemShut {NoStop}%
\end{thebibliography}%
\end{document}